\documentclass[%
 reprint,
 amsmath,amssymb,
 aps,
]{revtex4-2}

\usepackage{xcolor}
\usepackage{graphicx}
\usepackage{dcolumn}
\usepackage{bm}
\usepackage{hyperref}

\usepackage{xfrac}
\usepackage{esint}

\begin{document}

\preprint{APS/123-QED}

\title{Topology Optimization for Microwave Control With\\Reconfigurable Intelligent Metasurfaces In Complex Media}

\author{Theodosios D. Karamanos}
\email{theodosios.karamanos@espci.fr}
\author{Mathias Fink}
\author{Fabrice Lemoult}
\affiliation{Institut Langevin, ESPCI Paris, Université PSL, CNRS, 75005 Paris, France}

\begin{abstract}
\noindent Reconfigurable intelligent metasurfaces have been proposed as an efficient solution for improving wireless telecommunication systems in multiple scattering or reverberating media. Concurrently, topology optimization has been successfully employed as an inverse design technique in many fields, and particularly in electromagnetics. In this work, we apply a gradient-based topology optimization for tuning the binary elements of a metasurface for a focusing goal in a complex environment. First, the metasurface unit-cells are approximated as point sources and, then, the optimization problem is formulated. Afterwards, the proposed method is applied to find the optimal parameter sets on three distinct environments of an ascending complexity and the resulting focus for each case is demonstrated via simulations. The combination of reverberating cavity and a metasurface inside the latter reveals very powerful since everything can be solved analytically for focusing outside the cavity.
\end{abstract}


\maketitle
\section{Introduction}
Wireless telecommunications have experienced rapid growth in recent decades, approaching a threshold where the presence of multipath degrades signal quality. Over the years, the idea of smart electromagnetic environments has emerged, foreseeing a fully programmable wave propagation to harness this complexity and achieve optimized transmission of both information and power. 
Reconfigurable intelligent metasurfaces (RIS) have emerged as a promising technology in this direction, with applications in topics from efficient outdoor and indoor telecommunication and electromagnetic compatibility to imaging systems and quantum electrodynamics~\cite{di2020reconfigurable,tsilipakos2020toward,alexandropoulos2021reconfigurable}. 
Research on RIS has focused on various contemporary topics, including improved wireless communications~\cite{huang2019reconfigurable,basar2019wireless,elmossallamy2020reconfigurable}, indoor/cavity electromagnetic field shaping~\cite{kaina2014shaping,dupre2015wave,del2016spatiotemporal,gros2022multi} and metamaterial imaging~\cite{imani2020review,padilla2022imaging,saigre2022intelligent}. 
In principle, RIS are two-dimensional arrays consisting of subwavelength tunable elements, namely their electromagnetic response could be changed. Specifically, the unit-cells of a RIS could be individually modified to scatter the incident field by, for example, adding a phase. In the optical spectrum, this can be achieved by using the spatial light modulators (SLMs) as metasurface elements~\cite{popoff2010measuring,mosk2012controlling}. The SLMs consist often of liquid crystal cells which introduce a phase shift on the light they reflect (or pass through). The equivalent of the SLMs in the microwave regime are the spatial microwave modulators (SMMs)~\cite{kaina2014shaping,kaina2014hybridized}. In the aforementioned designs, an SMM consists of a rectangular static patch, as a main reflector, and a strip, as a parasitic resonator, which is binary tunable with the help of an embedded pin-diode. In particular, by controlling the bias of the diode, one can achieve a binary pixel in microwaves: it re-emits the incident wave with a $0$ or a $\pi$ phase shift, {\it i.e.} acting as a perfect electric or magnetic conductor. The SMMs provide an excellent, two-state unit-cell for reconfigurable metasurfaces in microwaves, although various other setups exist that could provide efficient alternatives~\cite{he2019tunable,beneck2021reconfigurable,ataloglou2023metasurfaces}. Another important theme in the research on RIS is their theoretical modelling, which could lead to efficient design and optimization. Contrary to common metasurface models~\cite{dimitriadis2015generalized,rahimzadegan2022comprehensive}, the unit-cells in RIS are not identical, thus, modifications to the usual models must be applied. Although circuit models are occasionally employed for the study of RIS~\cite{abeywickrama2020intelligent}, the use of microscopic models can increase the accuracy of the analysis, as well as provide a bigger understanding of the underlying physical problems~\cite{williams2020communication,danufane2021path,di2022communication}. Nevertheless, the lack or limited inclusion of mutual interaction between the unit-cells of the RIS or the inclusion of the coupling effects only on the phase of the scattered waves, may lead to inaccuracies and inefficient designs. 

Due to the dynamic nature of the communication systems or ever-changing goals require ``smarter", more efficient and faster methods for the reconfiguration of the tunable elements in comparison to ``brute-force" methods, currently employed in RIS~\cite{kaina2014shaping,dupre2015wave}. 
As a more efficient alternative, topology optimization is a promising tool for such applications~\cite{molesky2018inverse,christiansen2021inverse,li2022empowering, hammond2022high}. Originating from mechanical problems~\cite{bendsoe2003topology}, this density-based inverse design technique presents efficient devices for several purposes in electromagnetics, and, most notably, in photonics~\cite{molesky2018inverse,li2022empowering,augenstein2020inverse}. In this specific field, topology optimization has attracted considerable interest during the past years with examples including dielectric multiplexers~\cite{piggott2015inverse,su2018inverse}, metalenses~\cite{pestourie2018inverse,lin2019overlapping} and integrated photonic devices~\cite{augenstein2018inverse, augenstein2022inverse}. A particular characteristic of the topology optimization is that it is a gradient-based optimization technique, where the gradient is, usually, acquired efficiently and swiftly using the adjoint method~\cite{molesky2018inverse, christiansen2021inverse}. 
Topology optimization as an inverse design technique allows for a point-by-point material density variation in the structure under study, while it has demonstrated that it can produce optimized solutions for problems with hundreds of thousands of variables or more~\cite{molesky2018inverse, lin2019overlapping, hammond2022high}. These features of topology optimization are very promising for potential applications in RIS, where thousands of tunable elements may need to be reconfigured fast to achieve pre-determined, yet changing, goals.

In this work, we develop a topology optimization technique to be used for binary reconfigurable intelligent metasurfaces~\cite{kaina2014shaping,kaina2014hybridized}, when placed in complex media. 
Our framework is based on a combination of the existing gradient-based topology optimization techniques, with the use of the adjoint method, and the modelling of electromagnetic problems with Green's functions. 
First, we theoretically formulate the electromagnetic problem in 2D by assuming that the reconfigurable elements of the metasurface are two-state, {\it i.e.} $0$ or $\pi$ phase shift, and that they can be equivalently represented as point-scatterers. Moreover, the optimization problem is formulated with the goal of maximizing the intensity at a given focus point via a certain binary phase set of the metasurface elements.
Then, the adjoint method is employed to ensure the fast computing of the gradients of the objective function. Eventually, the optimization problem is reformulated using the 2D Green functions for the representation of the interactions between the elements of the metasurface, the focus and the source, ensuring an even faster overall optimization process. Afterwards, we apply the developed theoretical formulation to focusing problems in three different environments of an ascending complexity. We examine the efficiency of the resulting focus after the application of the featured topology optimization scheme using different options for the interactions between the unit-cells of the metasurface, while the possibility of the analytical calculation or numerical extraction of the 2D Green function values is explored for each complex medium under study. Notably we show that placing the source and the metasurface in a partially opened-cavity is a promising configuration since every thing can be solved analytically.
\section{Theoretical formulation}
\subsection{Problem Model}
\begin{figure}[t]
\centering 
\includegraphics[width=0.35\textwidth]{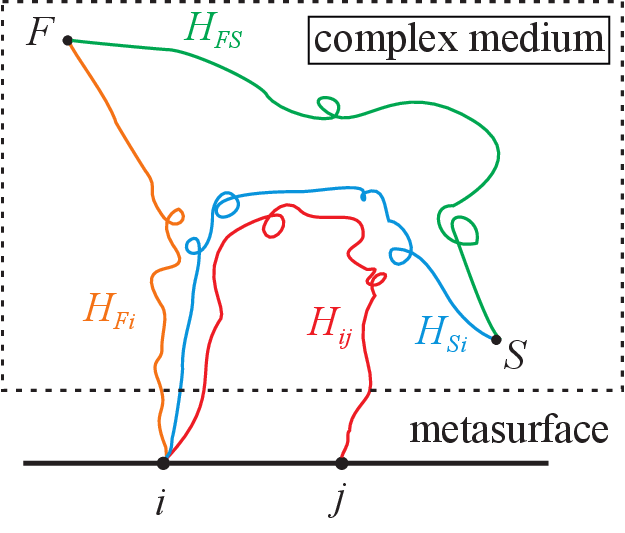}\vspace{-0mm}
\caption{Depiction of the generalized 2D focusing problem in a complex medium and in the presence of a reconfigurable metasurface.} \label{fig:mts-setup}
\end{figure}
Let us first model the problem of using a reconfigurable metasurface in a complex environment, as depicted in Fig.~\ref{fig:mts-setup}, in a generalized fashion for both 3D and 2D setups. Examples of complex media are highly-reflecting or multiple scattering environments~\cite{moustakas2000communication}, {\it e.g.} an urban landscape, an office or domestic room with furnitures, a cluster of dielectric or metallic scatterers, or a leaky closed cavity. The goal of the overall setup is set on electromagnetic focusing at a given point $F$ the energy coming from one or several emitters by the help of a tunable metasurface. Without loss of generality, we  assume, here, that the system is excited only by a single fixed point current source $\textbf{J}_{S}$ at the point $S$ (Fig.~\ref{fig:mts-setup}).
The reconfigurable metasurface consists of $N$ subwavelength elements positioned on a lattice with a unit-cell dimension $d < \lambda$. Moreover, in this work, the metasurface is considered as made of elements that totally reflect the incident wave (whatever the angle) but the phase $\phi$ of reflection can be tuned from $0$ to $\pi$. These two-state unit cells are ideal versions of a common theme with nowadays technology for reconfigurable metasurfaces~\cite{kaina2014shaping}. 
Mathematically, the reflectivity of pixel $i$ can be expressed as $R_i = e^{-\mathrm{i}\phi_i}$ with $\phi_i \in \{0,\pi\}$.

In that context we decide to model the metasurface elements as current point sources at the center of the unit-cells. To model the desired reflectivity, the re-emitted electric field by a point-like pixel $i$ corresponds to $\mathbf{b}_i = R_i\mathbf{E}_i^{\rm \,loc} = R_i(-\mathrm{i}\omega\mu  \mathbf{J}_i)$, where $\mathbf{E}_i^{\rm \,loc}$ is the local field at the element position to be retrieved, and $\mathbf{J}_i$, the equivalent current source. This approximation is valid as long as the dimensions of scatterers are much smaller than the operational wavelength and in the far field or radiation zone~\cite{jackson1999classical}.
It should be noted that both $R_i$ and $\mathbf{E}_i^{\rm \,loc}$ depend on $\phi_i$. Overall, the electromagnetic problem of a metasurface in a complex medium can be expressed using the Helmholtz equation as:
\begin{equation}\label{helmholtz-equation}
\left( \nabla^2 + \frac{\omega^2}{c^2(\mathbf{r})} \right)\textbf{E} =  \mathbf{b}_s\,\delta(\mathbf{r} - \mathbf{r}_S) + \sum_{i=1}^N \mathbf{b}_i(\phi_i)\, \delta(\mathbf{r} - \mathbf{r}_i)    
\end{equation}
\noindent with $\mathbf{b}_s = -\mathrm{i}\omega\mu\mathbf{J}_{\rm S}$ corresponding to the electric field generated by the source $S$. It should be noted that the complex medium of Fig.~\ref{fig:mts-setup} is non-homogeneous, thus, the wave velocity in \eqref{helmholtz-equation} is a function of the spatial position.

The optimization problem, in this case, consists in finding which binary phase configuration of the $N$ elements will result to the largest intensity value at the target point $F$ (Fig.~\ref{fig:mts-setup}). It corresponds to maximizing the figure of merit (FOM) $M = \|\mathbf{E}_{F}\|^2$.
In order to efficiently and rapidly find all the phases $\phi_i$ that will maximize the FOM, in this work, we will use the topology optimization method for inverse design.
\subsection{Topology Optimization and Adjoint Method}
In general, for structural, as well as electromagnetic topology optimization, a density-based material parametrization is used~\cite{christiansen2021inverse}. The material is represented by the continuous design field $p_i \in [0,1]$, which maps the material distribution for each pixel $i$ of the design domain. Therefore, the electromagnetic optimization problem is formulated, with the use of \eqref{helmholtz-equation} as:
\begin{subequations}\label{optimization-problem}
\begin{align}
\max_{\mathbf{p}}:&\,\,\, M = \|\mathbf{E}_{F}\left(p_i\right)\|^2 \\
s.t.:& \,\,\,\left( \nabla^2 + \frac{\omega^2}{c^2(\mathbf{r})} \right)\textbf{E} = \notag \\
&=\mathbf{b}_s \,\delta(\mathbf{r} - \mathbf{r}_S)  + \sum_{i=1}^N \mathbf{b}_i(p_i)\,\delta(\mathbf{r} - \mathbf{r}_i)  \\
&\text{with}\quad 0 \leq p_i \leq 1,\quad i = \{1, 2, ... \,,N\}.\nonumber
\end{align}\vspace{-0mm}
\end{subequations}

For the case of the tunable metasurface, 
the ``material" is, herein, the phase shift induced by the pixel of the metasurface. However, since only two states are allowed, {\it i.e.} $0$ or $\pi$, a binarization scheme needs to be applied on the variables $p_i$. 
To that end, we apply a \textit{smoothed Heaviside function}: 
\begin{equation}\label{heaviside-binarization}
\tilde{p}_i(p_i) = \frac{{\rm tanh}\left(\frac{\beta}{2}\right) + {\rm tanh}\left(\frac{p_i - \beta}{2}\right)}{{\rm tanh}\left(\frac{\beta}{2}\right) + {\rm tanh}\left(\frac{1 - \beta}{2}\right)},\quad \beta>1,
\end{equation}
where $\beta$ is a threshold value.
After binarization, a \textit{linear interpolation} is applied, resulting to the phase of element $i$ as:
\begin{equation}\label{linear-interpolation}
\phi_i(\tilde{p}_i) = \pi\,\tilde{p}_i - \mathrm{i}\alpha\,\tilde{p}_i\,(1 - \tilde{p}_i).
\end{equation}
The parameter $\alpha$ has, here, the role of a second binarization scheme; it controls the non-physical imaginary term that introduces an attenuation for all intermediate values between $0$ and $\pi$~\cite{christiansen2021inverse,hammond2022high}. Finally, in order to ensure an almost perfect binary result for the phase values, we impose a continuation scheme by gradually increasing $\beta$, with a ratio $\beta_{\rm inc}$ for each individual run of the optimization algorithm. The iterations and, eventually, the whole optimization process stop when the \textit{grey indicator}~\cite{augenstein2022inverse} is almost zero, or:
\begin{equation}\label{gray-indicator}
\gamma = \frac{1}{N}{\sum_{i=0}^{N}} 4p_i\,(1-p_i) \approx 0,
\end{equation}
In this way, the resulting  vector $\mathbf{p}$ is almost always of value $0$ or $1$. As an initial set of values, we use $p_i = 0.5$ for all $i$. It should be noted that the values of the parameters $\alpha$, $\beta$ and $\beta_{\rm inc}$ are problem dependant and a trial-and-error process is required to identify the best ones. Generally, low values for $\alpha$, $\beta$ and $\beta_{\rm inc}$ could provide better focusing results at the cost of longer optimization algorithm running times, effectively a trade-off process.

Afterwards, we use a gradient-based algorithm to solve the topology optimization problem above. This specific family of techniques utilizes the gradients, or sensitivities, $\frac{\mathrm{d}M}{\mathrm{d}\mathbf{p}}$. One could approximate the gradients via finite differences, but this process would involve solving or simulating the problem described by~\eqref{optimization-problem} for every combination of the design variables $p_i$, leading to too high computational times. 
Therefore, in this work, we utilize the more commonly used \textit{adjoint sensitivity analysis} or, simply, the \textit{adjoint method}~\cite{johnson2012notes,molesky2018inverse,christiansen2021inverse,luce2023merging}.  The strategy requires to solve only two distinct problems in order to compute all the required gradients. First, the solution of the so-called "direct" problem gives the field $\mathbf{E}_F$, and thus its intensity, at the focusing point $F$ when $S$ is emitting. Second, an adjoint problem is solved where $F$ becomes the source with an amplitude given by $\frac{\mathrm{d}M}{\mathrm{d}\textbf{E}}\Big|_{{\rm F},\,\mathbf{p}} = {\rm Re}\left\{E_F(\mathbf{p})\right\} - \mathrm{i}~{\rm Im}\left\{E_F(\mathbf{p})\right\}$. In particular, it permits to evaluate the adjoint solution $\mathbf{E}_{\rm adj}$ which corresponds to the E-field values at the $N$ metasurface elements. 
Hence, the gradients are obtained as (see \textit{Supplementary Material}):
\begin{gather}\label{adjoint-analysis}
\frac{\mathrm{d}M}{\mathrm{d}\mathbf{p}} = ~2\,{\rm Re}\Big\{\mathbf{E}_{\rm adj}\cdot\frac{\partial \mathbf{b}}{\partial \boldsymbol{\Phi}} 
\frac{\partial \boldsymbol{\Phi}}{\partial \mathbf{\tilde{p}}}\frac{\partial \mathbf{\tilde{p}}}{\partial \mathbf{p}}\Big\}\nonumber
\\ \\
\text{with}\quad \left( \nabla^2 + \frac{\omega^2}{c^2(\mathbf{r})} \right)\mathbf{E}_{\rm adj} = \frac{\mathrm{d}M}{\mathrm{d}\mathbf{E}}\Big|_{{F},\,\mathbf{p}}\,\delta(\mathbf{r}-\mathbf{r}_F) .\nonumber
\end{gather}
%
After the required derivative $\frac{dM}{d\textbf{p}}$ is obtained via the adjoint method, it is inserted in the optimizer \texttt{fmincon}, that performs nonlinear constrained optimization, and is included in MATLAB$^{\rm TM}$ 2021b~\cite{OptimizationToolbox}.

In the presented problem of a metasurface in a complex medium, $\mathbf{E}_{\rm adj}$ can be acquired, at least through numerical simulations, via two steps, a direct calculation and an adjoint one, while the terms $\frac{\partial \boldsymbol{\Phi}}{\partial \mathbf{\tilde{p}}}$ and $\frac{\partial \mathbf{\tilde{p}}}{\partial \mathbf{p}}$ are calculated analytically (see \textit{Supplementary Material}). 
The term $\frac{\partial \mathbf{b}}{\partial \boldsymbol{\Phi}}$ is not as easily retrievable as the other ones via the model introduced in \eqref{helmholtz-equation}, due to the complex interaction between the elements of the metasurface through the complex medium. However, if one assumes that the elements do not interact with each other, and the local fields on the metasurface pixels depend only on the source and the medium, {\it i.e.} they do not depend on the phases of the individual elements, then, $\frac{\partial b_i }{\partial \phi_i} = -\mathrm{i}e^{-\mathrm{i}\phi_i}\mathbf{E}_i^{\rm \,loc}$. Afterwards, the local fields at the centers of the unit cells can be calculated during the direct simulation. This approximation is quite common in reconfigurable metasurface applications and has been known to provide adequate results. Nevertheless, the omission of the dependence of the field due to the other elements  introduces inaccuracies and inefficiencies, which become larger when reconfigurable metasurfaces are placed in media apart from free space. Therefore, to alleviate these issues, in this work, we introduce a rigorous Green function model, which leads to accurate calculations and potentially smaller optimization times.
\subsection{Green function formulation}
The gradient in  \eqref{adjoint-analysis} can be retrieved through the calculation of $\mathbf{E}_{F}$. This can be achieved with the use of Green functions. 
In order to demonstrate this approach, let us, at this point of the analysis and without loss of generality, transform the general three-dimensional electromagnetic problem of \eqref{helmholtz-equation} in two dimensions. In order to do so, we assume that the source at the point $S$ is an out-of-plane line current (or towards the $z$-axis for a $xy$-plane), $\textbf{J}_S = I_S \hat{\textbf{z}}$, while the equivalent sources at the unit-cell centers are similarly induced out-of-plane line currents, $\textbf{J}_i = I_i \hat{\textbf{z}}$. Moreover, all materials involved are assumed to be invariant in the out of plane direction. Therefore, all E-field involved in this 2D problem only have a $\hat{\textbf{\textbf{z}}}$-component. At this point, we use the Green function $H_{\boldsymbol{\rho}\boldsymbol{\rho}'}$ 
as the electric field value, $E_z$, measured at coordinate $\boldsymbol{\rho}$ when a  point source is situated at location $\boldsymbol{\rho}'$ (see \textit{Appendix A} for freespace solutions).
The positions $\boldsymbol{\rho}$ and $\boldsymbol{\rho}'$ could be replaced by the positions of the source $S$, the focal spot $F$ or the pixel elements $i$, as shown in Fig.~\ref{fig:mts-setup}. Since the Helmholtz equation is reciprocal, even in an inhomogeneous environment, it holds that $H_{\boldsymbol{\rho}\boldsymbol{\rho}'} = H_{\boldsymbol{\rho}'\!\boldsymbol{\rho}}$.

By the help of these Green functions, we can fully calculate $E_{F}$. In a matrix form it writes:
\begin{equation}
\label{ER-calculation-a}
E_{\rm F} = H_{F\!S}\,b_s + \mathbf{H}^{\rm T}_{F}\,\mathbf{b},   
\end{equation}
\noindent where $\mathbf{H}^{\rm T}_{F} = [H_{{F}1}\,H_{{F}2},... H_{{F}\!N}]$, contains the Green function values between the focusing point $F$ and the metasurface elements and $\mathbf{H}^{\rm T}_{S} = [H_{{S}1}\,H_{{S}2}, ... H_{{S}\!N}]$ between the source point $S$ and the metasurface elements. 

In order to take into account the multiple interactions between the pixels of the metasurface, we need to  rewrite the field on the pixels as:
\begin{equation}
\label{ER-calculation-b}
\mathbf{b} = [R]\Big\{\sum_{n=0}^{+\infty}\big([H][R]\big)^n\Big\}
\mathbf{H}_{ S}b_s
\end{equation}
The square matrix $[H]$ contains the inter-elements coupling, that is to say the Green functions $H_{ij}$, and $[R] = {\rm diag}\{R_1\,R_2, ... R_N\}$ stands for the reflection coefficients of each metasurface pixel. The $n \in \mathbb{N}$ number in \eqref{ER-calculation-b} refers to the number of reflections between the environment and the metasurface that are taken into account for the calculation. If the problem involves a low reflecting medium, {\it e.g.} a leaky room with a few objects inside~\cite{kaina2014shaping}, then no or single reflection ($n=0$ or $n=1$) may be a good approximation. 
For the general case of infinite reflections, the power series involves square matrices and convergence actually always holds for passive systems. Therefore, the magic of this matrix formulation is that one can readily write the final solution of the multiply scattering series as~\cite{lax1952multiple}:
\begin{equation}\label{ER-calculation-c}
\mathbf{b} = [R]\Big([I] - [H][R]\Big)^{-1}\mathbf{H}_{S}b_s,   
\end{equation}
\noindent with $[I]$ being the identity matrix. Hence, through \eqref{ER-calculation-c}, the term  $\frac{{\rm d}\mathbf{b}}{{\rm d}\boldsymbol{\Phi}}$ in \eqref{adjoint-analysis} can now be analytically calculated for and infinite number of interactions between unit cells (see \textit{Supplementary Material}).

It is evident, that if we know all the Green function values for \eqref{ER-calculation-a},\eqref{ER-calculation-b} and \eqref{ER-calculation-c}, we can, afterward, find the desired gradient from \eqref{adjoint-analysis} and, finally, begin the optimization solver for the specific problem.
These Green function matrix or vector elements can either be extracted via numerical simulations or measurements in a pre-optimization step or be analytically calculated for certain environments. 
\section{Focusing applications}
\begin{figure*}[ht!]
\centering 
\includegraphics[width=0.475\textwidth]{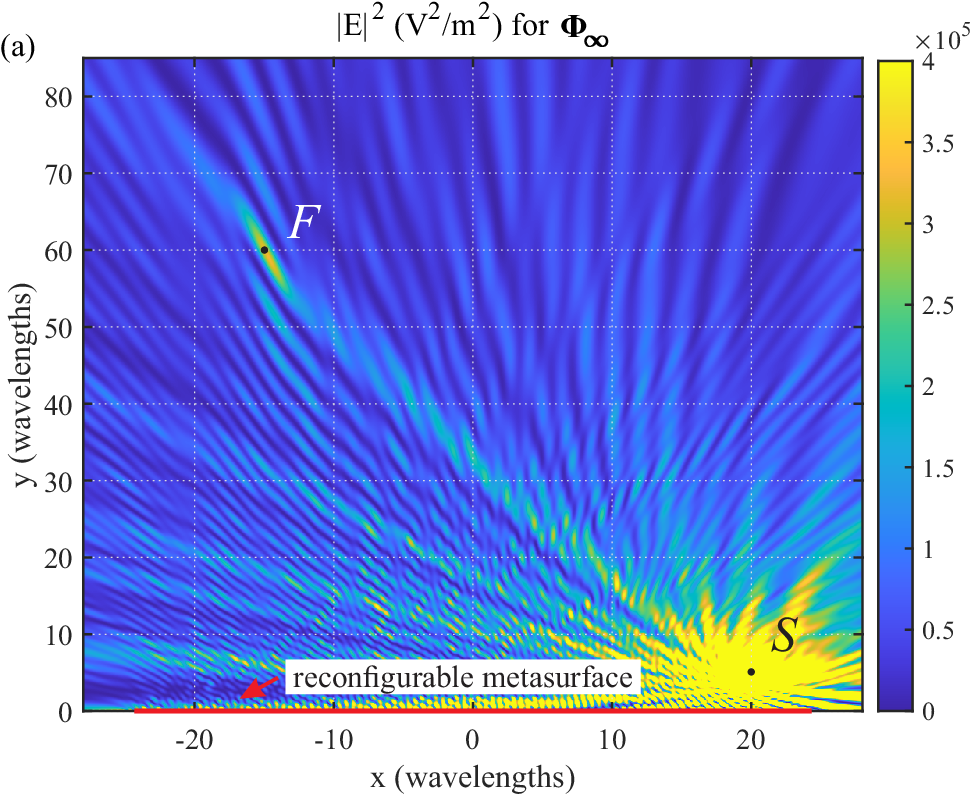}\qquad
\includegraphics[width=0.475\textwidth]{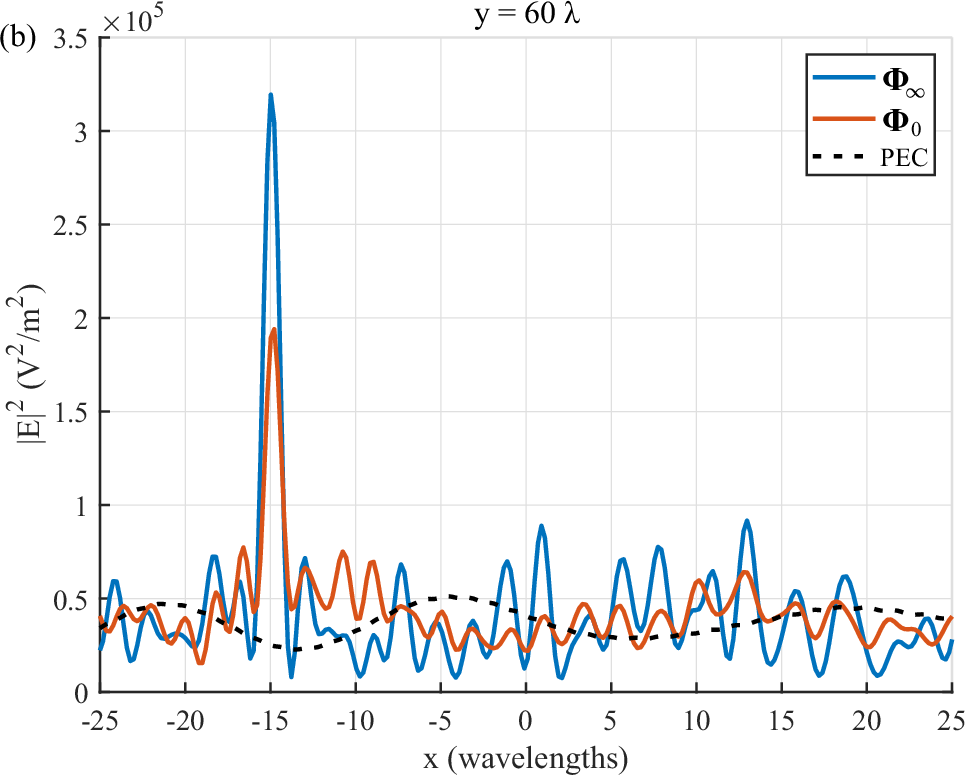} \vspace{-0mm}
\caption{Open space problem; (a) Mapping of the intensity values via COMSOL$^{\rm TM}$ simulation,
with the phase set $\boldsymbol{\Phi}_{\infty}$ as input. (b) The E-field intensity values for $y = 60\lambda$ and the various phase sets  obtained by the proposed topology optimization process.
\vspace{-0mm}} \label{fig:free-space}
\end{figure*}
After presenting the proposed topology optimization technique in the previous section, we  now apply it on various $2$D  focusing problems of a reconfigurable metasurface in a complex medium. First, we apply the proposed technique for a tunable metasurface in open space, where the Green functions required are replaced by the analytical free-space 2D ones. Next, we optimize the focusing problem for a reconfigurable metasurface placed inside a complex envrionment mimicking an office room, which takes the form of a leaky cavity. In this case, the Green functions are not known, therefore, they are extracted via numerical simulations before the optimization technique is applied. This acquisition step is time consuming but once all the Green functions are known the optimization runs quickly. Finally, we address the problem of a leaky-cavity antenna, where the metasurface and the source are placed in a partially open reverberating cavity. For this specific geometry, the 2D Green functions can be analytically calculated. 

Whenever the Green functions are needed to be extracted or a visualization of the results is required, we utilize COMSOL Multiphysics$^{\rm TM}$\cite{comsol} as a simulation tool. In particular, each example is reconstructed in COMSOL in the manner of Fig.~\ref{fig:mts-setup}, and the equivalent sources replacing the metasurface unit cells are set up as a function of the retrieved phases, where $b_i(\phi_i) = R_i(\phi_i) \textbf{E}_i^{\rm loc}$, as explained in Section II.A above.

In all problems, we perform the required optimization exploring two cases: taking into account infinite or no interactions between the elements of the metasurface. The latter is the most common in metasurface models~\cite{dupre2015wave, abeywickrama2020intelligent}, due to the simplicity it offers for modeling. Nevertheless, it is expected to produce worse maxima for the optimization problem in comparison with the full interaction model, especially as the environment complexity increases. Infinite interactions are considered if one uses~\eqref{ER-calculation-c}, while no interaction is included if \eqref{ER-calculation-b} is used with $n=0$. 
At the end, the resulting phase sets, $\boldsymbol{\Phi}_{\infty}$ and $\boldsymbol{\Phi}_{0}$, for respectively the infinite and no interaction cases are inserted into the full model of~\eqref{ER-calculation-a} and are compared for their intensity outputs. 

Finally, in this paper, we choose to work in the microwave spectrum in which many applications on wireless communications, involving reconfigurable metasurfaces are employed, both outdoors and indoors~\cite{di2020reconfigurable, tsilipakos2020toward, kaina2014shaping, dupre2015wave}. Thus, the operational frequency is set to $f = 2.4$~GHz, while the distance between the elements of the metasurface is chosen as $d=\lambda/2$, in order to have Shannon sampling of the metasurface~\cite{dimitriadis2015generalized,rahimzadegan2022comprehensive}. Nevertheless, the procedure presented is general, and can be utilized for smaller element distances, as well as other frequencies and applications, {\it e.g.} imaging with light using SLMs~\cite{mosk2012controlling}.
\subsection{Open space}
\begin{figure*}[ht!]
\centering 
\includegraphics[width=0.45\textwidth]{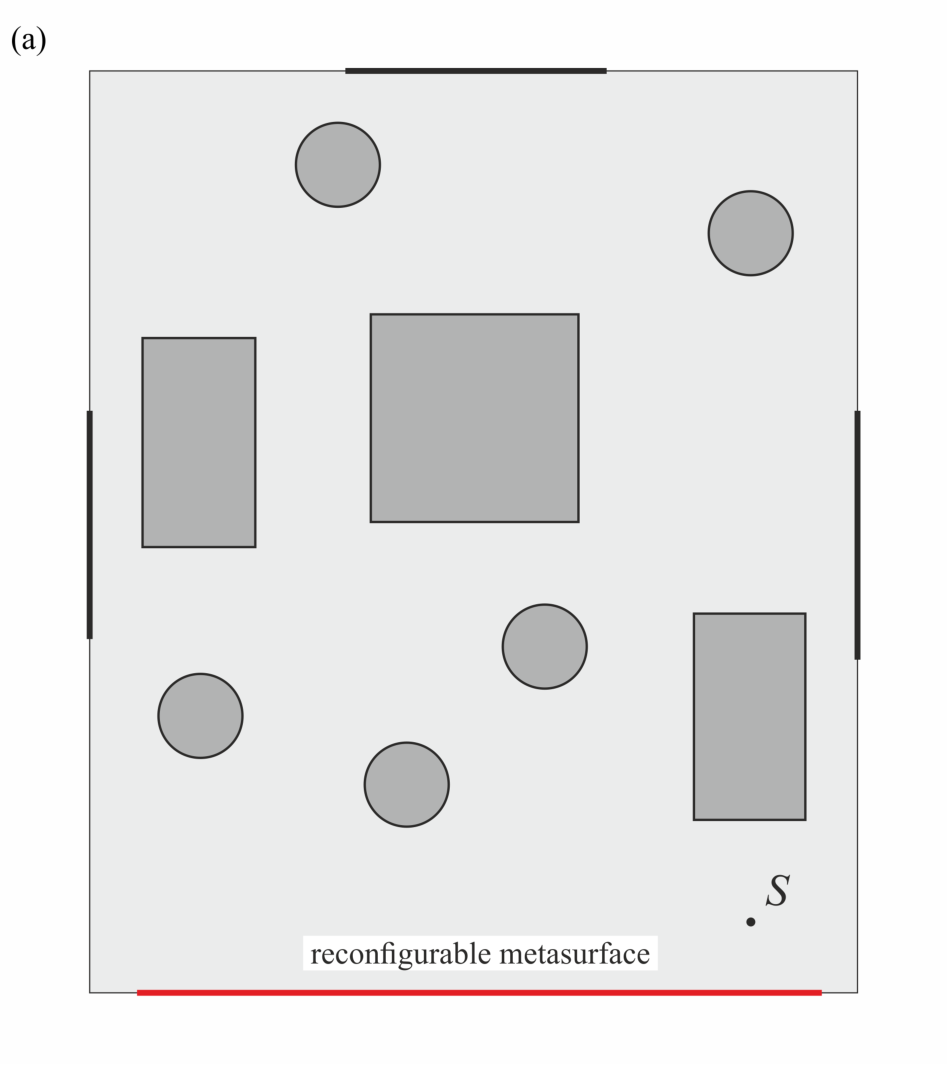}
\qquad
\includegraphics[width=0.45\textwidth]{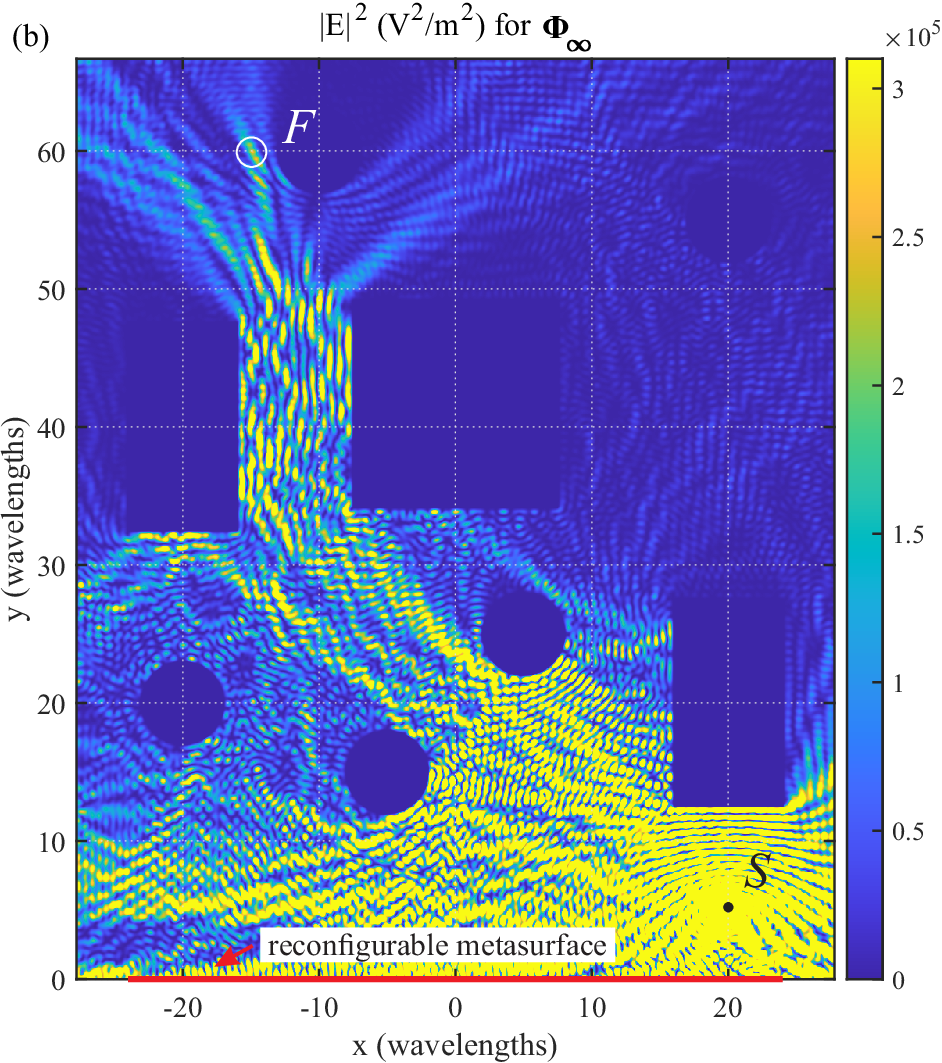} \vspace{-0mm}
\caption{Complex leaky cavity problem; (a) Illustration of the leaky cavity/room: The circular and square obstacle have their surfaces made of PEC. The boundaries of the cavity are constructed via the ``transition boundary condition" of COMSOL$^{\rm TM}$, with the bold lines representing ``windows" with material parameters $\varepsilon_{\rm r} = 1.5$, $\mu_{\rm r} = 1$, $\sigma = 0$ and with thickness, $th = \lambda$, and with the normal lines representing ``walls" with material parameters $\varepsilon_{\rm r} = 5$, $\mu_{\rm r} = 1$, $\sigma = 0$ and with thickness, $th = 2\lambda$. The dimensions of the cavity are set to $W\,{\rm x}\,H = 1.1Nd \,{\rm x}\, 1.32Nd$. (b) Mapping of the intensity values via COMSOL$^{\rm TM}$ simulation,
with the phase set $\boldsymbol{\Phi}_{\infty}$ as input.
\vspace{-0mm}} \label{fig:leaky-cavity-1}
\end{figure*}
Let us, first, apply the proposed optimization technique to the simplest case for the setup illustrated in Fig.~\ref{fig:mts-setup} where the ``complex medium" is replaced by free-open space. This specific case enables a fast and easy analytic approach. First, the Green functions, $H_{{F}i}$ and $H_{{S}i}$ required for the calculations in~\eqref{ER-calculation-a}, have an analytical expression~\cite{volakis2012integral,jackson1999classical} (see \textit{Appendix A}). Second, in the absence of a complex, reflective or absorbing environment, only the direct interaction between the metasurface elements exists, thus, the $H_{ij}$ are also the analytical textbook's formulas that only depend on the distance between $i$ and~$j$.

In this example, we employ a metasurface of $N=101$ reconfigurable elements; the center of the metasurface is placed at the origin of the axes, $O(0,0)$. Moreover, we place a point source with $I_{S} = 1$~A at the point $S(20\lambda,5\lambda)$. The optimization goal is, then, arbitrary set to the maximization of the intensity at the point $F(-15\lambda,60\lambda)$, as formulated in~\eqref{optimization-problem}.
The values of the binarization parameters are set to $\alpha = 30$, $\beta = 2$ and $\beta_{\rm inc} = 2$, as detailed in Section II.B.

The proposed topology optimization method is applied for this free-space example, for both cases of infinite and no interactions between the metasurface elements. After the desired phase set is analytically computed, the one from the infinite interaction analysis, $\boldsymbol{\Phi}_{\infty}$, is inserted in the COMSOL$^{\rm TM}$ simulation and the resulting E-field intensities are depicted in Fig.\ref{fig:free-space}(a). Nicely, a focus is achieved at the expected point $F$. 

Let us now compare the efficiency between the use of infinite and zero interaction, using the 2D Green functions for free space to analytically calculate the $|E|^2$ for $y = 60\lambda$ via~\eqref{ER-calculation-a}. 
Note that at this point the E-field is calculated by taking into account all the metasurface element interactions, for each phase set solution. Comparative results using $\boldsymbol{\Phi}_{\infty}$ and $\boldsymbol{\Phi}_{0}$ optimization solutions, as well as results using $\phi_i = \pi$ for each element, thus emulating a metallic reflector, are displayed in Fig.\ref{fig:free-space}(b). The final focus at $x = -15\lambda$ is clearly observed for both retrieved phase sets, however, a relative improvement of $70\%$ is calculated between zero and infinite interactions. Moreover, the relative improvement between the PEC reflector and the $\boldsymbol{\phi}_{\infty}$ solution lies at $\sim 1200\%$. The resulting phase sets from the optimization algorithm along with the comparison between the analytical intensity calculations and simulations are further provided in the \textit{Supplementary Material}.

The improvement in the final intensity at the focusing point using infinite interactions in comparison to zero is relatively small for a free-space problem but still non-negligible. Therefore, the approximation, commonly used in reconfigurable metasurfaces applications, of not considering it, is relatively efficient for open space setups. Nevertheless, as we will see in the following examples, highly reflective and complex environments deem the inclusion of the element interactions via the matrix $[H]$ in~\eqref{ER-calculation-b} necessary for large focusing effect in the optimization process. 
%
\subsection{Complex leaky cavity}
\begin{figure*}[ht!]
\centering 
\includegraphics[width=0.45\textwidth]{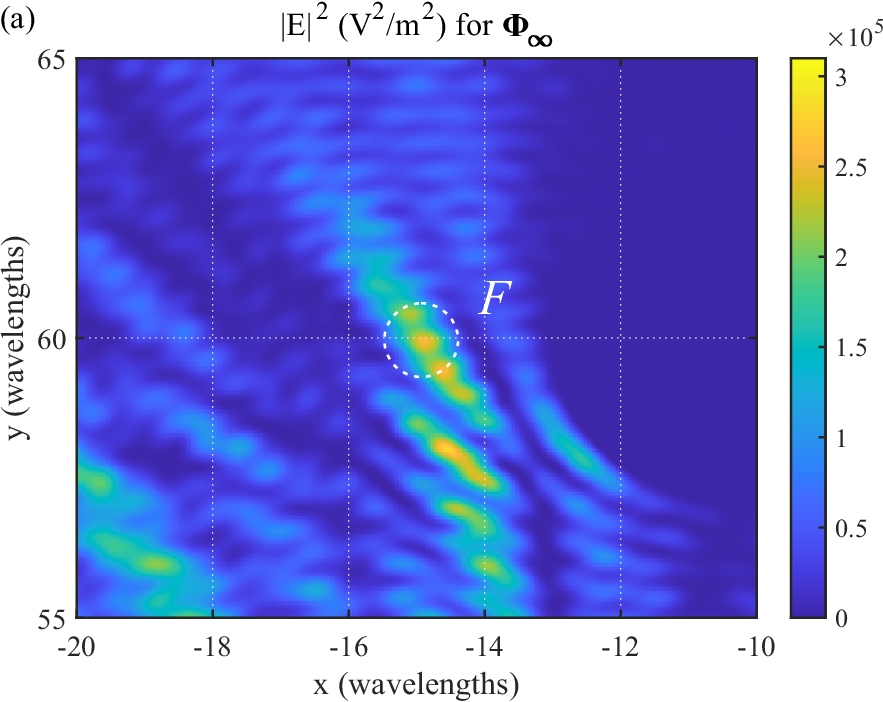}
\qquad
\includegraphics[width=0.45\textwidth]{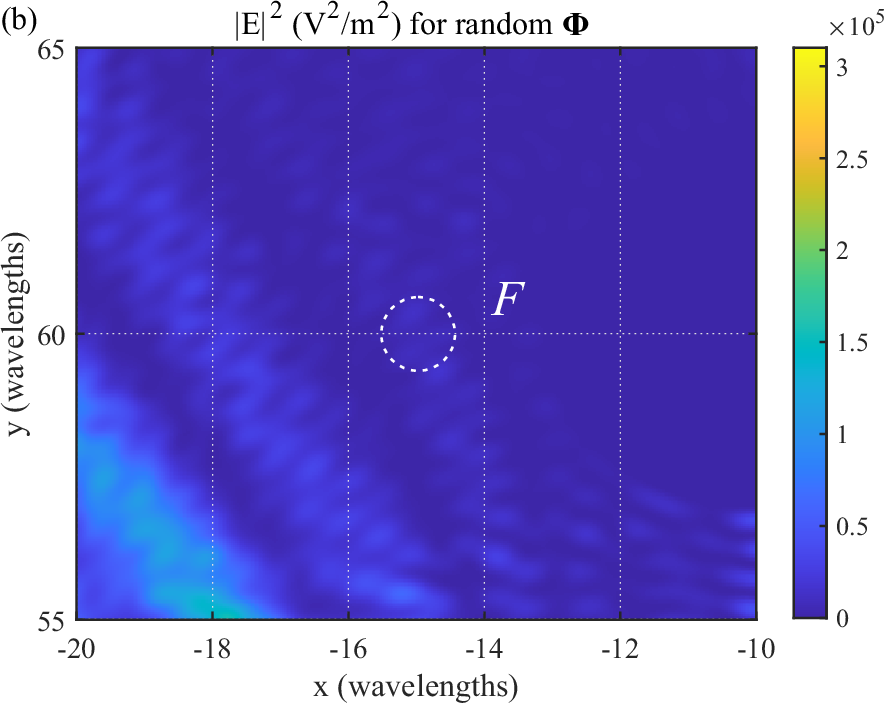} \vspace{-0mm}
\caption{Complex leaky cavity problem; Mapping detail of the intensity values via COMSOL$^{\rm TM}$ simulation around the focusing point (a) for a phase set $\boldsymbol{\phi}_{\infty}$ and (b) for a random phase set.
\vspace{-0mm}} \label{fig:leaky-cavity-2}
\end{figure*}
%
%
Next, the optimization algorithm is applied on a truly complex and random environment: a leaky room. This complex medium, depicted in Fig.~\ref{fig:leaky-cavity-1}(a), is modeled as a 2D orthogonal leaky cavity containing PEC obstacles of different shapes. Obviously, in this case, the 2D Green functions  cannot be obtained analytically. Thus, the 2D Green functions are extracted via simulations with COMSOL$^{\rm TM}$. Specifically, $H_{ij}$ are obtained by placing each time a point source with amplitude $I_i = 1$A at the position of element $i$ and measuring the E-field at the positions of the other $j$ elements, thus running $N$ independent simulations. From the same simulations, the values of $\textbf{H}_{F}$, $\textbf{H}_{S}$ are obtained in the meantime. For $H_{F\!S}$ an $(N+1)$th simulation is required. It should be noted that the values of $H_{ii}$ are not extracted, herein, and are approximated as zeros. This procedure of extracting 2D Green functions via simulations requires a lot of computational time, especially for a dense mesh. But, these values are characteristic of the problem's geometry and can be stored and used in future optimization runs. Specifically, if the source point and the metasurface position are fixed and the focus point is moving, like the case of stationary WiFi router in a room communicating with a moving device, the $[H]$ and $\textbf{H}_{S}$ values are already pre-extracted and stored and only a single simulation is required for the retrieval of $\textbf{H}_{F}$ and $H_{F\!S}$ as long as the rest of the environment does not change. Moreover, the same procedure of the 2D Green functions retrieval can be performed via measurements much faster, provided that the elements of the metasurfaces and the receiver at the focusing point have receiver/transmitter capabilities.

Again, as in the previous example, we use a metasurface of $N=101$ reconfigurable elements, placed at the origin of the axes. The source with $I_{S} = 1$~A is placed at $S(20\lambda,5\lambda)$, and the focusing point is at $F(-15\lambda,60\lambda)$, just like in the previous free-space problem.
The values of the binarization parameters are set to $\alpha = 40$, $\beta = 1.2$ and $\beta_{\rm inc} = 1.2$, as explained in Section II.B.

Once all 2D Green functions are extracted via simulations, the proposed topology optimization is, then, employed for both $n = 0$ and $n \rightarrow \infty$ element interactions in~\eqref{ER-calculation-b}. The resulting intensity inside the cavity for the optimal phase solutions for infinite interactions is depicted in Fig.\ref{fig:leaky-cavity-1}(b). $F$ appears to be on a region of increased intensity compared to neighbour points. The intensity around $F$ is also depicted in Fig.\ref{fig:leaky-cavity-2}(a) where the resulting focus is more clearly shown. In parallel, the absence of focusing around $F$ when a random set of phases is used is illustrated in Fig.\ref{fig:leaky-cavity-2}(b). The relative change of intensity between the random and the infinite interactions phase sets being roughly $\sim 4000\%$. Finally, the relative improvement for the intensity value between the zero and the infinite interactions solutions is of a factor $3.5$.
The resulting phase set solution from the topology optimization process, as well as the
intensity values along the lines $x=-15\lambda$ and $y=60\lambda$ are further provided in the \textit{Supplementary Material}.

In this complex environment, once the set of Green's functions is fully known the optimization procedure runs very quickly and finds a solution that effectively exhibits a focus near the target position $F$.  The overall improvement is far better than in the free-space thus making the use of RIS all the more interesting when used in complex and reverberating environment where each pixel provides an extra degree of freedom~\cite{lemoult2009manipulating,mosk2012controlling}. Note that the inclusion of infinite interactions between the
elements of the reconfigurable metasurface is mandatory as it has provided a solution that results in a considerably better focus in comparison with not including the interactions.
However, in this example, the improvement comes with a considerable cost of the computationally demanding pre-extraction of the matrix $[H]$. 

To tackle this issue, we propose to find a geometry of a complex medium, where the values of $H_{ij}$, $\textbf{H}_{F}$, $\textbf{H}_{S}$ and $H_{F\!S}$ can be pre-calculated analytically, thus providing speed in the topology optimization process. In the next example, we will examine one of such cases.
\subsection{Cavity antenna}
\begin{figure*}[ht!]
\centering 
\includegraphics[width=0.48\textwidth]{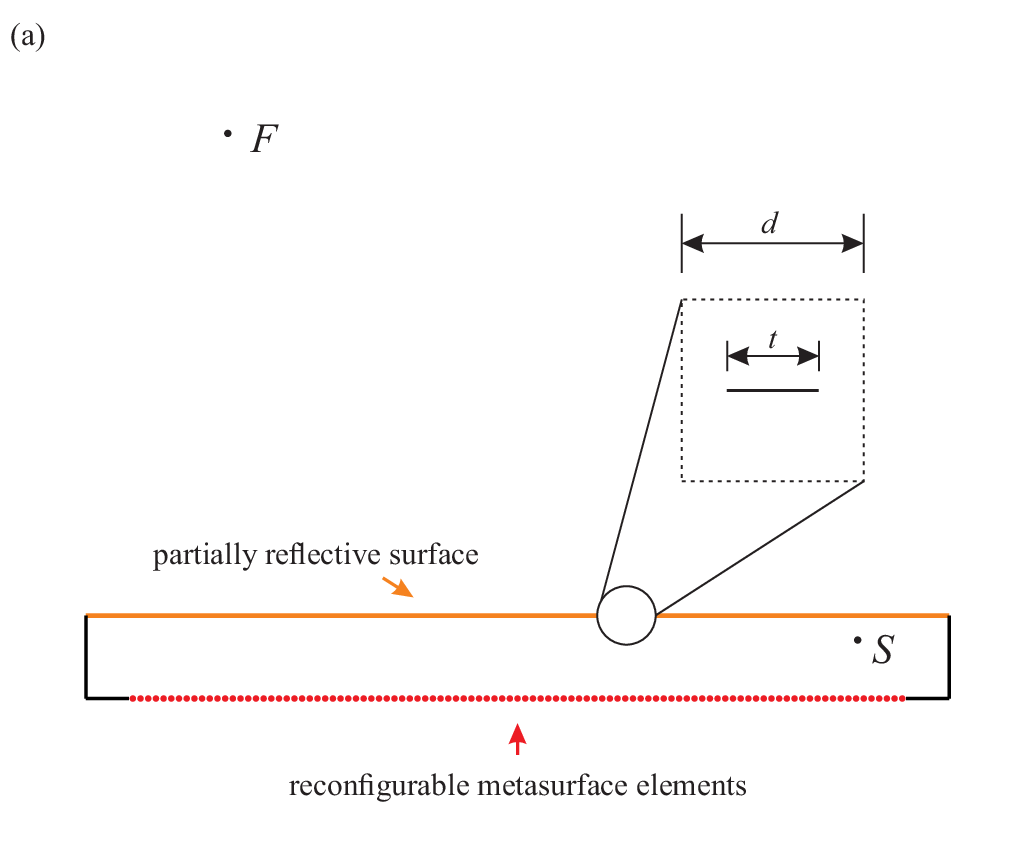}
\qquad
\includegraphics[width=0.47\textwidth]{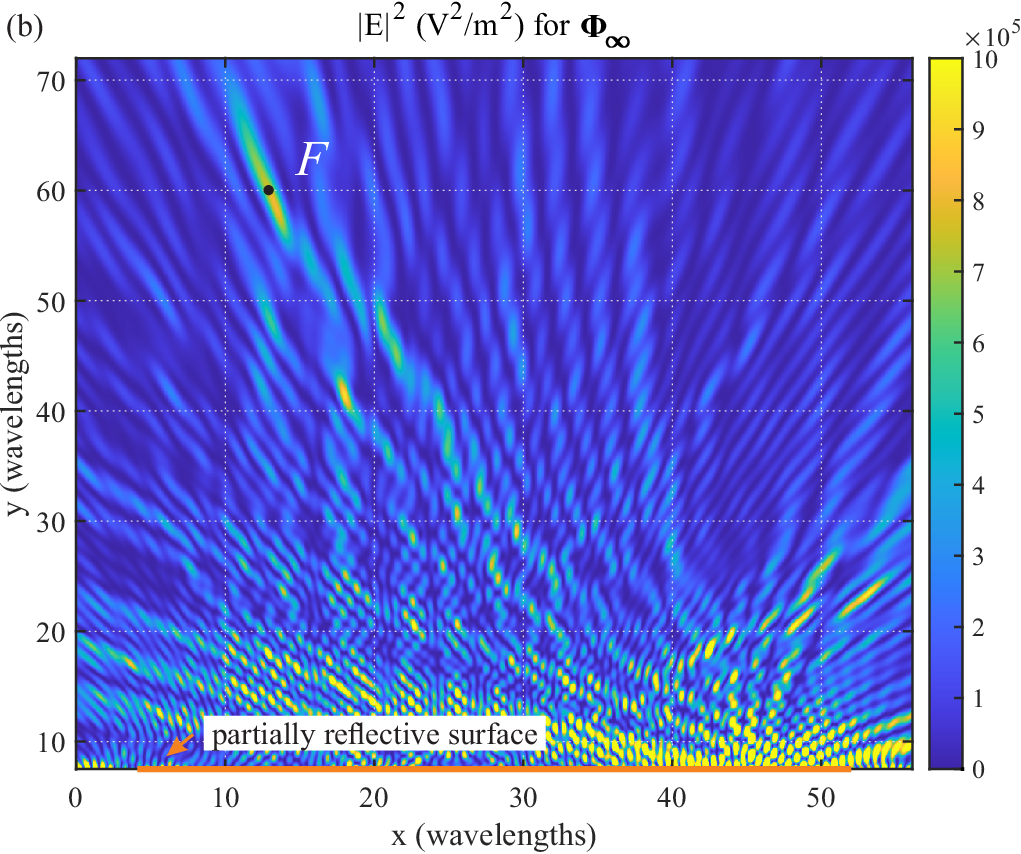} \vspace{-0mm}
\caption{Cavity antenna problem; (a) Depiction of the cavity antenna design: The height of the cavity is set to $h=7.5\lambda$, while its length is $L = 1.1(N+1)d$. Inlet setup; illustration of the unit cell of the partially reflective surface. The unit cell dimension is set to $d = \lambda/2$, while the dimension of the PEC strip placed is the middle of the unit-cell is set to $t=0.2d$. (b) Mapping of the intensity values via COMSOL$^{\rm TM}$ simulation,
with the phase set $\boldsymbol{\Phi}_{\infty}$ as input.
\vspace{-0mm}} \label{fig:2d-cavity-1}
\end{figure*}
\begin{figure}[ht]
\centering 
\includegraphics[width=0.47\textwidth]{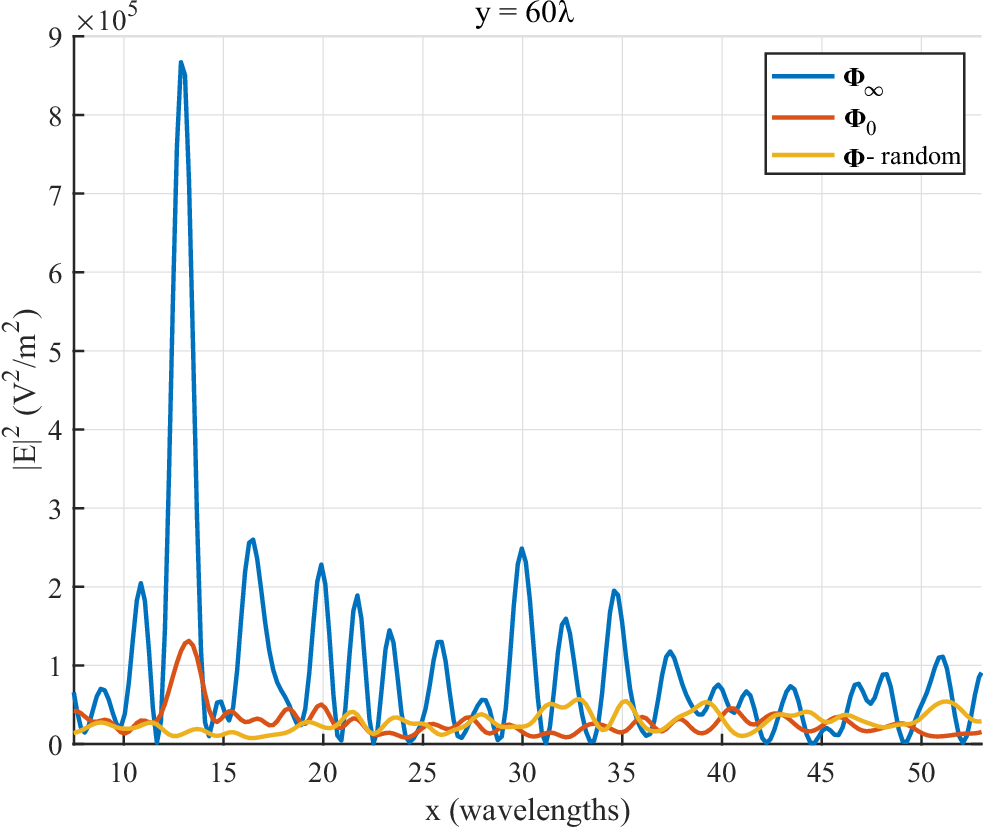}
\caption{Cavity antenna problem; The E-field intensity values for $y = 60\lambda$ and the various phase sets obtained by the proposed topology optimization process. \vspace{-0mm}} \label{fig:2d-cavity-2}
\end{figure}
For the last example, let us apply the proposed topology optimization algorithm on another complex and highly reflective environment, but where, this time, the necessary 2D Green functions are analytically calculated.
Such a geometry consists of a 2D cavity antenna as depicted in Fig.\ref{fig:2d-cavity-1}(a). The RIS is placed at the bottom of a reverberating cavity with the right and left walls of the cavity being made of perfectly conducting material. The fourth wall (top) is made of a partially reflective surface, which is composed of a subwavelength metallic grating (alternation of metal and free space); thus the cavity leaks to the surrounding environment. The source-feed is placed inside the cavity itself. The focusing point is for its part placed outside the 2D leaky cavity. This type of device constitutes a cavity antenna, used as an example in this subsection, and has attracted considerable attention both in academia and industry for applications involving, among others, satellite communications and radar~\cite{epstein2016cavity, gros2020tuning}.

The key point of this example is the analytical calculation of the 2D Green functions, which practically corresponds to the calculation of fields inside and outside of the cavity. First, the reflection/transmission coefficients from an infinite version of the partially reflective aperture are calculated for a plane wave incidence and an angle range $(-\pi,\pi)$. For the specific case of a 2D aperture array, the reflection/transmission coefficients can be analytically calculated~\cite{jackson1999classical}, or, they can be extracted via simulations, as we do specifically in this work. Afterwards, the calculation of the fields inside the cavity leads to the calculation of $H_{ij}$ and $H_{{S}i}$. In particular, for this purpose, we employ the \textit{method of images} to remove the walls of the cavities and calculate the fields inside via summations of 2D free space Green functions, taking into account as many images of the sources as required to achieve an adequate convergence. Finally, to evaluate the field outside the cavity the \textit{Kirchhoff's integral theorem} is used. From the fields on the partially reflective surface, the fields outside the cavity are obtained, and, thus, the required values of $H_{Fi}$ and $H_{F\!S}$. The formulas for the analytical calculation of the 2D Green functions for this cavity antenna example are provided in the~\textit{Appendix B}, while a more detailed analysis for their derivation is given in the~\textit{Supplementary Material}. 

Then, we again employ $N=101$ elements with $0$ or $\pi$ phase-states, and the left-bottom corner of the cavity is placed at $O(0,0)$, as depicted in Fig.\ref{fig:2d-cavity-1}(a). 
After placing a source at the point $S(48.05\lambda,5\lambda)$ and setting the optimization goal to focusing at the point $F(13.05\lambda,60\lambda)$, the proposed topology optimization technique is applied on the problem for both cases of infinite and zero element interactions. The binarization parameters are set to $\alpha = 40$, $\beta = 1$ and $\beta_{\rm inc} = 1.2$, as explained in Section II.B. The optimization is fully ran in Matlab with these analytical formulae, and then the optimal phase solutions are re-injected in COMSOL in order to visualise the field distribution outside the cavity. 

The mapping of the intensity values above the partially reflective surface for the resulting $\boldsymbol{\Phi}_{\infty}$ set is provided in Fig.\ref{fig:2d-cavity-1}(b). A focus is successfully created at the expected point $F$, confirming that both the analytical formulae and the topological optimization have worked.
Comparative results of $|E|^2$ at $y=60\lambda$ using the infinite  and zero interactions
optimization solutions, as well as results using a random pick of $0$ or $\pi$ phase values, are displayed in Fig.\ref{fig:2d-cavity-2}. The relative intensity change between the random phase set and the $\boldsymbol{\Phi}_{\infty}$ at the focus at $x=13.05\lambda$ is calculated to roughly $\sim 70000\%$, while the relative intensity change between the $\boldsymbol{\Phi}_{0}$ and the $\boldsymbol{\Phi}_{\infty}$ resulting sets is an improvement of more than $6$ times.
%
%
\subsection{Discussion}
%
%
From all the previous examples, we can first conclude that the topological optimization scheme manages to find optimal binary phase solutions that effectively create a focus. The focusing results are expected to improve with a larger number of $N$ reconfigurable elements. Also, as the complexity of the medium increases, the improvement in the resulting focusing becomes all the more remarkable. This is a direct consequence of the complexity of the medium that is turned onto an advantage by being able to control spatial degrees of freedom~\cite{lemoult2009manipulating,mosk2012controlling,kaina2014shaping}. The more complex the environment, the more useful the RIS. Or said differently, the more ``multipath" the different Green functions are, the more impact each individual pixel has. 

The second comment is a corollary of the previous one. As the Green function becomes more and more complex together with the propagating medium, there are more chances that waves are reflected back to the other pixels of the metasurface. Therefore a big difference is made in the algorithm on whether or not multiple interactions between the pixels are considered. Indeed, for the case of a metasurface in free-space the topology optimization results provide a barely better focus for infinite interactions  in comparison with zero. Therefore, the latter option is potentially viable for the common open-space, telecommunication examples~\cite{di2020reconfigurable}. However, it is later shown that the use of $n \rightarrow \infty$ in~\eqref{ER-calculation-b} instead of $n = 0$ provides visibly better results. Specifically, as the environment becomes increasingly more reflective and involves more modes~\cite{dupre2015wave},
the use of infinite interaction is essential for a obtaining a focus via the proposed topology optimization scheme. And interestingly, in terms of computational demands, it is more efficient since only a single matrix inversion is required as shown in~\eqref{ER-calculation-c}. 

Third, an important feature that greatly enhances the performance of the presented topology optimization technique is the prospect of the analytical calculation of the required 2D Green function values in~\eqref{ER-calculation-a} for certain geometries. In principal, the Green function values can always be extracted via simulations on a pre-optimization step, as demonstrated in the leaky room example of Section III.B. In practice, though, this requires a lot of computational time and resources, and even if the values of $[H]$ and $\textbf{H}_{S}$ characterize the geometry of a time-invariant complex medium, and can be stored and reused for multiple optimization runs, one must still extract each time the values of $\textbf{H}_{F}$ and $H_{F\!S}$ for a changing position of the desired focusing point $F$. However, this problem is alleviated if the geometry of the complex medium permits an analytical calculation of the Green function, as performed in the example of Section III.C. This not only obviously enables a much faster overall optimization process, but also enables a real-time experimental use of topology optimization, particularly useful for telecommunication applications, where the focusing point/receiver constantly changes position~\cite{dupre2015wave, gros2020tuning}. The confirmation of the analytical solution with simulation has been performed here, but undoubtedly the next step will be to test this scenario experimentally. 

Finally, although the analysis presented herein is formulated and performed for two-dimensional problems, it is general in nature and can be expanded, with some care, to three-dimensional complex media. Apart from the different types of Green functions that must be used (see \textit{Appendix A}), one has to, also, take into account for the 3D cases, the different polarizations, the different multipole types (electric and magnetic) and the choice of the multipolar order. These considerations will eventually produce more complicated matrices and vectors in the respective~\eqref{ER-calculation-a} for 3D environments but the formalism should remain valid. 
\section{Conclusion}
In this paper, we have presented a topology optimization technique to perform electromagnetic focusing, when binary reconfigurable metasurfaces are utilized in complex media. First, the optimization problem was formulated in 2D, with the elements of the metasurface approximated as point-sources and with the use of Green functions. Moreover, the adjoint method was employed for the fast retrieval of the necessary derivative. 
Subsequently, the developed method was applied on various focusing examples in complex environments, for 2D Green function values analytically calculated or retrieved via simulations. The obtained phase values of the reconfigurable elements provided excellent focusing at the intended points, while it was also demonstrated the consideration of infinite interactions between metasurface elements in the topology optimization scheme provides significantly large intensity values at the focus. 

Considering future work, we aim to equivalently expand the proposed topology optimization technique on 3D problems involving complex media, as well as to experimentally use the presented technique to real-time open space or cavity problems.
\begin{acknowledgments}
T.K. would like to thank Yannick Augenstein for the long and fruitful discussions on the theory and algorithms of topology optimization. We thank Steven Johnson for initiating us to the vast world of topology optimization. This work has received support under the program ``Investissements d’Avenir'' launched by the French Government, from the Simons Foundation/Collaboration on Symmetry-Driven Extreme Wave Phenomena, and from the "Agence Innovation Defense" under the RAPID m3SFA project.
\end{acknowledgments}
{\appendix
\section{Two-dimensional Green’s function for free space}
Let us consider a 2D space, described by the cylindrical coordinate system $(\rho,\theta)$ and with the vector to an observation point, $\boldsymbol{\rho} = \rho\boldsymbol{\hat{\rho}} =  x\mathbf{\hat{x}} + y\mathbf{\hat{y}} = \rho \left({\rm cos}\theta \mathbf{\hat{x}} +  {\rm sin}\theta \mathbf{\hat{y}}\right)$. A unitary point source is placed at $\boldsymbol{\rho}'$ with an imposed current along direction $\hat{\textbf{z}}$. Additionally, the medium is inhomogeneous, yet, it does not affect the polarization of the propagating waves. Hence, the problem is described by the Helmholtz equation:
\begin{equation}\label{2d-green-function-free-1}
\left(\nabla^2 + \frac{\omega^2}{c^2(\boldsymbol{\rho})} \right)H(\boldsymbol{\rho},\boldsymbol{\rho}')\hat{\textbf{z}} = -\delta(\boldsymbol{\rho} - \boldsymbol{\rho}')\hat{\textbf{z}}.
\end{equation}
The solution to this problem $H(\boldsymbol{\rho},\boldsymbol{\rho}')$ is called the Green function. For the case of the unbounded, homogeneous \textit{free space} it is the perpendicular two-dimensional Green’s function~\cite{volakis2012integral},
\begin{equation}\label{2d-green-function-free-2}
H(\boldsymbol{\rho},\boldsymbol{\rho}') = G_{\rm 2d} (\|\boldsymbol{\rho}-\boldsymbol{\rho}'\|) =  -\frac{\mathrm{i}}{4}H^{(2)}_0(kR),
\end{equation}
where $H^{(2)}_0(.)$ denotes the zeroth-order Hankel function of the second kind, $k=\frac{\omega}{c}$ the freespace wavenumber, and $R=\|\boldsymbol{\rho} - \boldsymbol{\rho}'\|$.
\section{Calculation for the $[H]$, $\textbf{H}_{ F}$, $\textbf{H}_{ S}$ and $H_{ FS}$ values for cavity antenna problem}
The problem under study is the leaky cavity antenna depicted in \ref{fig:2d-cavity-1}(a) and the application of the proposed topology optimization technique requires the calculation of the Green function values in \eqref{ER-calculation-a} and ~\eqref{ER-calculation-c}. The reflection and transmission coefficients at the partially reflective surface, or simply $r$ and $t$, are a function of the angle $\theta$ of the incident wave and can be either calculated or extracted via simulations.  

To emulate the presence of the walls the method of images is applied~\cite{jackson1999classical,volakis2012integral}. It should be noted that the metasurface elements are placed  exactly at the bottom of the cavity and therefore there is no reflection on this wall. Then, the $H_{ij}$ and $H_{Si}$ can be calculated analytically. Therefore, $H_{ij}$ is calculated as,
\begin{subequations}\label{Hij}  
\begin{equation}
H^{\,\rm d}_{ij} = \sum_{-m}^m \Big\{G_{\rm 2d}(d^{\,+}_{ij,m}) - G_{\rm 2d}(d^{\,-}_{ij,m}) \Big\},\quad i \neq j    
\end{equation}\vspace{0mm}
\begin{equation}
\begin{split}  
H^{\,\rm r}_{ij} = \sum_{-m}^m \Big\{r(\theta^{\,+}_{ij,m})&\,G_{\rm 2d}(q^{\,+}_{ij,m}) \\
- r&(\theta^{\,+}_{ij,m})\,G_{\rm 2d}(q^{\,-}_{ij,m}) \Big\}, \quad i \neq j   
\end{split}   
\end{equation}\vspace{-0mm}
\begin{equation}    
H_{ij} = H^{\,\rm d}_{ij} + H^{\,\rm r}_{ij},
\end{equation}
\end{subequations}
where $d_{ij,m}^{\,\pm} = |x_j - (\pm x_i + 2mL)|$, $\theta_{ij,m}^{\,\pm} = {\rm tan}^{-1}(d_{ij,m}^{\,\pm}/2h)$ and $q^{\,\pm}_{ij,m} = \sqrt{4h^2 + (d_{ij,m}^{\,\pm})^2}$ with $m \in \mathbb{Z}$, according to Fig.\ref{fig:2d-cavity-1}(a). The values $x_i$ and $x_j$ correspond to the x-coordinates of the unit-cells centres. Moreover, the value $|m|$ indicates the number of images considered in the analysis. Given the fact that $G_{{\rm 2d}}$ stands from a cylindrical wave the further the source the lower the magnitude of the Green's function, a value of $|m| = 20-30$ provides a very good accuracy. 

The elements of the main diagonal of $[H]$ are themselves calculated as:
\begin{equation}\label{Hii}
\begin{split} 
H_{ii} = r(0)&\,G_{\rm 2d}(2h)  + \\
\sum_{-m}^m &\Big\{G_{\rm 2d}(d^{\,+}_{ii,m}) - G_{\rm 2d}(d^{\,-}_{ii,m}) \Big\}.  
\end{split} 
\end{equation}
The $H_{Si}$ are similarly calculated as,
\begin{subequations}\label{His}  
\begin{equation}
H^{\,\rm d}_{Si} = \sum_{-m}^m \Big\{G_{\rm 2d}(q^{\,+}_{1,i,m}) - G_{\rm 2d}(q^{\,-}_{1,i,m}) \Big\},   
\end{equation}\vspace{0mm}
\begin{equation}
\begin{split}  
H^{\,\rm r}_{Si} = \sum_{-m}^m \Big\{r(\theta^{\,+}_{Si,m})\,G_{\rm 2d}&(q^{\,+}_{2,i,m}) \\
- r&(\theta^{\,+}_{Si,m})\,G_{\rm 2d}(k\,q^{\,-}_{2,i,m}) \Big\},
\end{split}   
\end{equation}\vspace{-0mm}
\begin{equation}    
H_{Si} = H^{\,\rm d}_{Si} + H^{\,\rm r}_{Si},
\end{equation}
\end{subequations}
\noindent where $d_{Si,m}^{\,\pm} = |2mL \pm x_{\rm S} - x_i|$, $\theta_{Si,m}^{\,\pm} = {\rm tan}^{-1}(d_{Si,m}^{\,\pm}/(2h-y_{\rm S}))$, $q^{\,\pm}_{1,i,m} = \sqrt{y_{\rm S}^2 + (d_{Si,m}^{\,\pm})^2}$ and $q^{\,\pm}_{2,i,m} = \sqrt{(2h^2-y_{\rm S})^2 + (d_{Si,m}^{\,\pm})^2}$  with $m \in \mathbb{Z}$.

Afterwards, the calculation of the remaining values of $H_{Fi}$ and $H_{F\!S}$ requires the accurate calculation of the fields outside the cavity. This is possible by utilizing the \textit{Kirchhoff integral} \cite{jackson1999classical}, where all infinitesimally section of the partially reflective surface of the Section III.C problem are considered point sources. Therefore, the integral is transformed for the current problem of Fig.\ref{fig:2d-cavity-1}(a) as,
\begin{equation}\label{kirchhoff-int}
\begin{split}
E(\boldsymbol{\rho}_{F}) = \int_0^L& E(\boldsymbol{\rho}') \Big(\mathbf{\hat{y}}\cdot\nabla'G_{\rm 2d}\left(\|\boldsymbol{\rho}_{\!F}-\boldsymbol{\rho}'\|\right)\Big) \\
&+\, G_{\rm 2d}\left(\|\boldsymbol{\rho}_{\!F}-\boldsymbol{\rho}'\|\right)\Big(\mathbf{\hat{y}}\cdot\nabla'E(\boldsymbol{\rho}') \Big) {\rm d}x,    
\end{split}
\end{equation}
where the vector $\boldsymbol{\rho}'$ represents the position of each piece of the integral, while $\boldsymbol{\rho}_{\!F}$ represents the position of the point where the field is to be calculated, herein, the focusing point. Hence, $\|\boldsymbol{\rho}_{\!F}-\boldsymbol{\rho}'\| = R_{F} = \sqrt{(x_{F} - x')^2 + (y_{F} - h)^2}$. The right-hand part of~\eqref{kirchhoff-int} is reformulated as,
\begin{equation}\label{kirchhoff-int-2}
\begin{split}
\mathbf{\hat{y}}\cdot\nabla'G_{\rm 2d}(R_{F}) = \qquad \qquad \qquad \,\,\, &\\
=\frac{|y_{\rm F} - h|}{\sqrt{(x_{\rm F} - x')^2 - (y_{\rm F} - h)^2}} &\Big[\frac{\mathrm{i}k}{4}H^{(2)}_1(kR_{F})\Big].    
\end{split}
\end{equation}

Then, the problem of calculating $H_{Fi}$ and $H_{F\!S}$ essentially becomes a problem of calculating the E-fields and their derivatives on each point of the partially reflected surface. Specifically, the E-field at $F$ will be calculated using \eqref{kirchhoff-int} after placing a current point source at the metasurface elements positions or at $S$ and using the method of images. Therefore, for each metasurface element $i$ it is derived that,
\begin{subequations}\label{Ei}  
\begin{equation}
\begin{split} 
E_{\,ix'} = \sum_{-m}^m \Big\{t(&\theta^{\,+}_{ix',m})\,G_{\rm 2d}(u^{\,+}_{i,m}) \\
-t&(\theta^{\,-}_{ix',m})\,G_{\rm 2d}(u^{\,-}_{i,m}) \Big\}\,(-\mathrm{i}\omega\mu I_i),  
\end{split} 
\end{equation}
\begin{equation}
\begin{split}  
\mathbf{\hat{y}}\hspace{-0.5mm}\cdot\hspace{-0.5mm}\nabla'&E_{\,ix'} = \frac{-\mathrm{i}k}{4}\sum_{-m}^m\Big\{\frac{h}{u^{\,+}_{i,m}}
t(\theta^{\,+}_{ix',m})H^{(2)}_1(k\,u^{\,+}_{i,m}) \\-
&\frac{h}{u^{\,-}_{i,m}}
t(\theta^{\,-}_{ix',m})H^{(2)}_1(k\,u^{\,-}_{i,m})
\Big\}\,(-\mathrm{i}\omega\mu I_i),
\end{split}   
\end{equation}
\end{subequations}
\noindent where $u_{i,m}^{\,\pm} = \sqrt{h^2 + (x' + 2mL \pm x_{i})^2}$ and $\theta_{ix',m}^{\,\pm} = {\rm tan}^{-1}(u_{i,m}^{\,\pm}/h)$  with $m \in \mathbb{Z}$. Similarly, for the source point $S$, it holds that,
\begin{subequations}\label{Es}  
\begin{equation}
\begin{split} 
E_{\,{S}x'} = \sum_{-m}^m \Big\{t(&\theta^{\,+}_{{S}x',m})\,G_{\rm 2d}(u^{\,+}_{{S},m}) \,- \\
t&(\theta^{\,-}_{{S}x',m})\,G_{\rm 2d}(u^{\,-}_{{S},m}) \Big\}\,(-\mathrm{i}\omega\mu I_i),  
\end{split} 
\end{equation}
\begin{equation}
\begin{split}  
\mathbf{\hat{y}}&\hspace{-0.5mm}\cdot\hspace{-0.5mm}\nabla'E_{\,{S}x'} \hspace{-0.5mm}= \hspace{-0.5mm}\frac{-\mathrm{i}k}{4}\sum_{-m}^m\hspace{-0.5mm}\Big\{\frac{h}{u^{\,+}_{{\rm S},m}}
t(\theta^{\,+}_{{S}x',m})H^{(2)}_1(ku^{\,+}_{{\rm S},m}) \\
& -\frac{h}{u^{\,-}_{{\rm S},m}}
t(\theta^{\,-}_{{S}x',m})H^{(2)}_1(ku^{\,-}_{{\rm S},m})
\Big\}\,(-\mathrm{i}\omega\mu I_i),
\end{split}   
\end{equation}
\end{subequations}
\noindent where $u_{{S},m}^{\,\pm} = \sqrt{(y_{S} - h)^2 + (x' + 2mL \pm x_{S})^2}$ and $\theta_{{S}x',m}^{\,\pm} = {\rm tan}^{-1}(u_{{S},m}^{\,\pm}/(y_{S} - h))$  with $m \in \mathbb{Z}$. After that, inserting~\eqref{kirchhoff-int-2} and~\eqref{Ei} into~\eqref{kirchhoff-int} produces $E_{Fi}$, which in turn leads to the calculation of the vector element $H_{Fi} = E_{Fi}/(-\mathrm{i}\omega\mu I_i)$. Finally, inserting~\eqref{kirchhoff-int-2} and~\eqref{Es} into~\eqref{kirchhoff-int} gives $E_{{F\!S}}$ with the last value required for~\eqref{ER-calculation-a}, $H_{F\!S} = E_{{F\!S}}/(-\mathrm{i}\omega\mu I_i)$.
}

\clearpage

%
%
\bibliographystyle{apsrev4-2}
%

\newpage
\onecolumngrid
\setcounter{page}{1}
\setcounter{section}{0}

\renewcommand{\theequation}{S\arabic{equation}} 
\renewcommand\thefigure{S\arabic{figure}}  
\renewcommand\refname{References and Notes}

\vspace{2mm}
\begin{center}
{\LARGE Supplementary Material for \vspace{2mm} \\ 
``Topology Optimization for Microwave Control With \vspace{2mm} \\ Reconfigurable Intelligent Metasurfaces In Complex Media''}  
\end{center}\vspace{1mm}

\begin{center}
Theodosios D. Karamanos$^{\ast}$, Mathias Fink and Fabrice Lemoult
\\
Institut Langevin, ESPCI Paris, Université PSL, CNRS, 75005 Paris, France
\\
$^\ast$e-mail: theodosios.karamanos@espci.fr.
\end{center}

\section{Application of the adjoint method in Section II}
Consider the generic optimization problem of the minimization of an objective function $M$, relative to a field $\textbf{x}$, which in turn is a function of a set of controllable parameters $\textbf{g}$. The field $\textbf{x}$ is subject of a partial differential equation (PDE). Thus, the general optimization can be reformulated as,
\begin{subequations}\label{s-optimization-problem-1}
\begin{align}
\min_{\mathbf{g}}:&\,\,\, M\big(\mathbf{x}(\mathbf{g})\big) \\
subject\,\, to:& \,\,\,\bar{\bar{A}}(\mathbf{g})\,\mathbf{x}(\mathbf{g}) = \mathbf{b}(\mathbf{g}) 
\end{align}\vspace{-0mm}
\end{subequations}
where $\bar{\bar{A}}\mathbf{x} = \mathbf{b}$ expresses the discretization of the PDE, with
$\bar{\bar{A}} \in \mathbb{R}^{\,n {\rm x} n}$ is the system matrix, $\mathbf{x} \in \mathbb{R}^{\,n}$ is the solution field vector and $\mathbf{b} \in \mathbb{R}^{\,n}$ are the sources. In gradient-based optimization, the retrieval of the gradient of the objective function with respect to the design variables $\mathbf{g}$. The first step towards this  is to use the chain rule as,
\begin{equation}\label{s-chain-rule}
\frac{{\rm d}M}{{\rm d}\textbf{g}} = \frac{{\rm d}M}{{\rm d}\textbf{x}}\frac{{\rm d}\textbf{x}}{{\rm d}\textbf{g}}.
\end{equation}
While the term $\frac{{\rm d}M}{{\rm d}\textbf{x}}$ can usually be calculated analytically, the term $\frac{{\rm d}\textbf{x}}{{\rm d}\textbf{g}}$ is more difficult to acquire. If we use the PDE, then,
\begin{equation}\label{s-dxdg-1}
\textbf{x} = \bar{\bar{A}}^{-1}\mathbf{b} \Rightarrow \frac{{\rm d}\textbf{x}}{{\rm d}g_i} = \frac{{\rm d}\bar{\bar{A}}^{-1}}{{\rm d}g_i}\mathbf{b} +  \bar{\bar{A}}^{-1}\frac{{\rm d}\mathbf{b}}{dg_i} =
\bar{\bar{A}}^{-1} \left(\frac{{\rm d}\mathbf{b}}{{\rm d}g_i} -  \frac{{\rm d}\bar{\bar{A}}}{{\rm d}g_i}\textbf{x}\right) ,
\end{equation}
and, if we combine \eqref{s-chain-rule} with \eqref{s-dxdg-1}, we arrive to
\begin{equation}\label{s-dxdg-2}
\frac{{\rm d}M}{{\rm d}\textbf{g}} = \frac{{\rm d}M}{{\rm d}\textbf{x}} \bar{\bar{A}}^{-1}\,\left(\bigg[\frac{{\rm d}\mathbf{b}}{{\rm d}g_1},\frac{{\rm d}\mathbf{b}}{{\rm d}g_2},\dots,\frac{{\rm d}\mathbf{b}}{{\rm d}g_N}\bigg] - \bigg[\frac{{\rm d}\bar{\bar{A}}}{{\rm d}g_1}\textbf{x},\frac{{\rm d}\bar{\bar{A}}}{{\rm d}g_2}\textbf{x},\dots,\frac{{\rm d}\bar{\bar{A}}}{{\rm d}g_N}\textbf{x}\bigg]\right) = 
\frac{{\rm d}M}{{\rm d}\textbf{x}} \bar{\bar{A}}^{-1}\,\left(\frac{{\rm d}\mathbf{b}}{{\rm d}\textbf{g}} - \frac{{\rm d}\bar{\bar{A}}}{{\rm d}\textbf{g}}\textbf{x}\right),
\end{equation}
where we assumed an $N$ number of design variables $g$.

It is very inefficient to solve the system \eqref{s-dxdg-2} of so many equations. Specifically, to acquire $\frac{{\rm d}M}{{\rm d}\textbf{g}}$ for all design variable will require $N$ matrix inversions.
In order to retrieve the gradient easier, we use the \textit{adjoint method}~\cite{johnson2012notes-s, molesky2018inverse-s}. Let us define the adjoint solution as,
\begin{equation}\label{ajoint-def}
\textbf{x}_{\rm adj} = \frac{{\rm d}M}{{\rm d}\textbf{x}} \bar{\bar{A}}^{-1},
\end{equation}
which is the solution of \textit{the adjoint problem} \cite{molesky2018inverse-s},
\begin{equation}\label{ajoint-problem-real}
\bar{\bar{A}}^{\,\dagger}\,\textbf{x}_{\rm adj} = \frac{{\rm d}M}{{\rm d}\textbf{x}} ,
\end{equation}
where $\dagger$ is the adjoint operator. 
Then, \eqref{s-dxdg-2} becomes: 
\begin{equation}\label{s-dxdg-3}
\frac{{\rm d}M}{{\rm d}\textbf{g}} =  
\textbf{x}_{\rm adj}\,\left(\frac{{\rm d}\mathbf{b}}{{\rm d}\textbf{g}} - \frac{{\rm d}\bar{\bar{A}}}{{\rm d}\textbf{g}}\textbf{x}\right).
\end{equation}
Therefore, $\frac{{\rm d}M}{{\rm d}\textbf{g}}$ can be acquired by solving the \textit{direct problem} for $\textbf{x}$ and the \textit{adjoint problem} for $\textbf{x}_{\rm adj}$, while $\frac{{\rm d}\mathbf{b}}{{\rm d}\textbf{g}}$ and $\frac{{\rm d}\bar{\bar{A}}}{{\rm d}\textbf{g}}$ can be usually calculated analytically. 

If $\bar{\bar{A}}$, $\mathbf{x}$ and $\mathbf{b}$ have complex components, which is the case in electromagnetic problems, the analysis is similar, with \eqref{s-dxdg-3} turning to:
\begin{equation}\label{s-dxdg-complex}
\frac{{\rm d}M}{d\textbf{g}} = 2\, {\rm Re} \Bigg\{ 
\textbf{x}_{\rm adj}\,\left(\frac{{\rm d}\mathbf{b}}{{\rm d}\textbf{g}} - \frac{{\rm d}\bar{\bar{A}}}{{\rm d}\textbf{g}}\textbf{x}\right)\Bigg\}.
\end{equation}
Furthermore, if the system is passive, it holds $\bar{\bar{A}}^{\dagger} = \bar{\bar{A}}$, hence, only two simulations/calculations of the PDE for different sources, can give fast the required gradient.

We now apply the adjoint method on the optimization problem presented in \textbf{Section II} in (5) of the \textit{Main document}, while the phase additions of the elements, $\boldsymbol{\Phi}$ are a function of the design variables $\tilde{\textbf{p}}(\textbf{p})$, as explained in the introduction of the binarization constraints in \textbf{Section II.B}. Thus, if we apply \eqref{s-dxdg-complex}, it becomes 
\begin{equation}\label{s-dxdg-complex-problem}
\frac{{\rm d}M}{{\rm d}\textbf{p}} = 2\, {\rm Re} \Bigg\{ 
\textbf{E}_{\rm adj}\,\frac{{\rm d}\mathbf{b}}{{\rm d}\mathbf{p}} \Bigg\} = 
2\, {\rm Re} \Bigg\{ 
\textbf{E}_{\rm adj}\,\left(\frac{{\rm d}\mathbf{b}}{{\rm d}\boldsymbol{\Phi}} \frac{{\rm d}\boldsymbol{\Phi}}{{\rm d}\tilde{\mathbf{p}}} \frac{{\rm d}\tilde{\mathbf{p}}}{{\rm d}\mathbf{p}} \right)\Bigg\}.
\end{equation}
where $\frac{{\rm d}\bar{\bar{A}}}{{\rm d}\textbf{p}} = 0$, because the complex environment of the problem does not depend on the changes of the reconfigurable elements of the metasurface.


Let us now apply the adjoint method on the problem expressed in \textbf{Section II} of the \textit{Main article}. From the definitions of the functions $\tilde{p}_i(p_i)$ and $\phi_i({\tilde{p}_i})$ in Section II.B, their respective derivatives can be analytically derived as,
\begin{subequations}\label{s-derivatives}
\begin{align}
\tilde{p}_i(p_i) = \frac{{\rm tanh}\left(\frac{\beta}{2}\right) + {\rm tanh}\left(\frac{p_i - \beta}{2}\right)}{{\rm tanh}\left(\frac{\beta}{2}\right) + {\rm tanh}\left(\frac{1 - \beta}{2}\right)}\quad &\Rightarrow \quad
\frac{{\rm d}\tilde{p}_i}{{\rm d}p_i} = \frac{\beta}{2}\,\frac{1 - {\rm tanh}^2\left(\frac{p_i\,\beta - \beta^2}{4}\right)}{{\rm tanh}\left(\frac{\beta}{2}\right) + {\rm tanh}\left(\frac{1 - \beta}{2}\right)},\quad \beta>1 \\
\phi_i(\tilde{p}_i) = \pi\,\tilde{p}_i - \mathrm{i}\alpha\,\tilde{p}_i\,(1 - \tilde{p}_i) \quad &\Rightarrow \quad
\frac{{\rm d}\phi_i}{{\rm d}\tilde{p}_i} = \pi -i\alpha(2\tilde{p}_i - 1).
\end{align}\vspace{-0mm}
\end{subequations}
Hence, \eqref{s-dxdg-complex-problem} becomes
\begin{equation}\label{s-dxdg-complex-problem-2}
\frac{{\rm d}M}{{\rm d}\textbf{p}} = 
2\pi\, {\rm Re} \Bigg\{ 
\textbf{E}_{\rm adj}\,\left(\frac{{\rm d}\mathbf{b}}{{\rm d}\boldsymbol{\Phi}} \frac{{\rm d}\tilde{\mathbf{p}}}{{\rm d}\mathbf{p}} \right)\Bigg\}.
\end{equation}
The calculation of $\frac{{\rm d}\textbf{b}}{{\rm d}\boldsymbol{\Phi}}$ is more complicated but can still be performed analytically via the~(9) presented in \textbf{Section II.C} for infinite interactions between the metasurface elements. Specifically, for each column of $\frac{{\rm d}\textbf{b}}{{\rm d}\boldsymbol{\Phi}}$ it holds that:
\begin{equation}\label{s-dbdphi-1}
\frac{{\rm d}\mathbf{b}}{{\rm d}\phi_i} = \frac{{\rm d}}{{\rm d}\mathbf{\phi}_i}\Bigg\{[R]\,\big( [I] - [H][R]\big)^{-1} \Bigg\}\,\textbf{H}_{S}b_s = \Bigg\{\frac{{\rm d}[R]}{{\rm d}\mathbf{\phi}_i}\,\big( [I] - [H][R]\big)^{-1} +
[R] \,\frac{{\rm d}}{{\rm d}\mathbf{\phi}_i}\big( [I] - [H][R]\big)^{-1} \Bigg\}\,\textbf{H}_{S}b_s.
\end{equation}
If the identity $\frac{{\rm d}K^{-1}}{{\rm d}x} = -K^{-1}\frac{{\rm d}K}{{\rm d}x}K^{-1}$ is used, the \eqref{s-dbdphi-1} finally becomes
\begin{equation}\label{s-dbdphi-2}
\frac{{\rm d}\mathbf{b}}{{\rm d}\phi_i} = \Bigg\{\frac{{\rm d}[R]}{{\rm d}\mathbf{\phi}_i}\,\big( [I] - [H][R]\big)^{-1} +
[R]\big( [I] - [H][R]\big)^{-1} [H]\frac{{\rm d}[R]}{{\rm d}\mathbf{\phi}_i}\big( [I] - [H][R]\big)^{-1} \Bigg\}\,\textbf{H}_{S}b_s,
\end{equation}
with $\frac{{\rm d}[R]}{{\rm d}\mathbf{\phi}_i}$ being a square $N\,{\rm x}\,N$ zero matrix matrix, with the exception of the $(i,i)$ element, $\frac{{\rm d}[R]}{{\rm d}\mathbf{\phi}_i}\big|_{ii} = -\mathrm{i}e^{-\mathrm{i}\phi_i}$. It should be noted that for zero interaction between metasurface elements considered, the calculation becomes much easier as,
\begin{equation}\label{s-dbdphi-0}
\frac{{\rm d}\mathbf{b}}{{\rm d}{\phi}_i} = \frac{{\rm d}[R]}{{\rm d}\mathbf{\phi}_i}\,\textbf{H}_{S}b_s.
\end{equation}
Therefore, if the Green function values $[H]$, $\textbf{H}_{F}$, $\textbf{H}_{S}$ and $H_{F\!S}$ are known either by analytical calculations or by extraction via simulations, the $\frac{{\rm d}\textbf{b}}{{\rm d}\boldsymbol{\Phi}}$ can be obtained.

Finally, the remaining term $\textbf{E}_{\rm adj}$ is calculated by solving the adjoint problem.
The adjoint solution is essentially the fields at the ``pixels" of the tunable medium, {\it i.e.} at the unit cells, and with a source with a value $\frac{{\rm d}M}{{\rm d}\textbf{E}}$ placed at the desired focusing point~\cite{molesky2018inverse-s,li2022empowering-s}. The value of the source for the adjoint problem is simply calculated as ~\cite{remmert1991theory-s}:
\begin{equation}\label{s-adjoint-source}
\frac{{\rm d}M}{{\rm d}\textbf{E}}\Big|_{{F},\mathbf{p}} = 
\frac{1}{2}\left(\frac{\partial M}{\partial ~{\rm Re}\{E_{F}(\mathbf{p})\}} - \mathrm{i}\,\frac{\partial M}{\partial~ {\rm Im}\{E_{F}(\mathbf{p})\}}\right)
= {\rm Re}\{E_{F}(\mathbf{p})\} -\mathrm{i}\,{\rm Im}\{E_{ F}(\mathbf{p})\},
\end{equation}
where $E_{\rm F}$ is the direct solution or the value of the E-field at the focusing point for the previous set of phases in the optimization process. Then, the  adjoint solution can be calculated as,
\begin{equation}\label{s-adjoint-solution}
\textbf{E}_{\rm adj} = \textbf{H}_{F}\,\frac{{\rm d}M}{{\rm d}E}\Big|_{{F},\mathbf{p}}\quad.
\end{equation}
Therefore, only two simulations or calculations are enough for the retrieval of all the necessary derivatives~$\frac{{\rm d}M}{{\rm d}\textbf{p}}$.

\section{Phase set solutions for the examples of Section III}
\begin{figure}[ht]
\centering 
\includegraphics[width=1\textwidth]{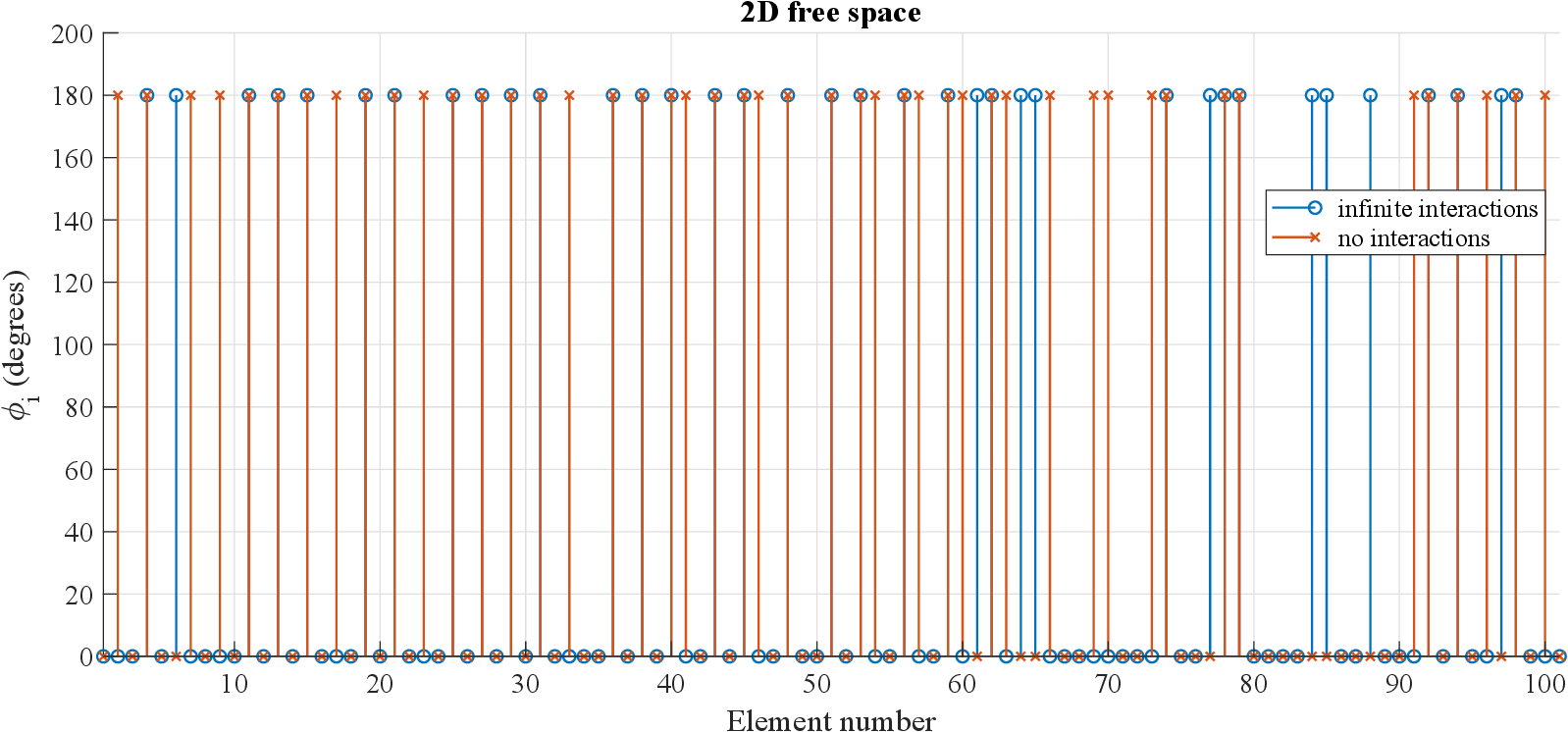} 
\caption{Phase set solutions, $\boldsymbol{\Phi}_{\infty}$ and $\boldsymbol{\Phi}_{0}$, for a RIS placed in 2D free space.
\vspace{1mm}} \label{fig:angles-1}
\end{figure}

\begin{figure}[!ht]
\centering 
\includegraphics[width=1\textwidth]{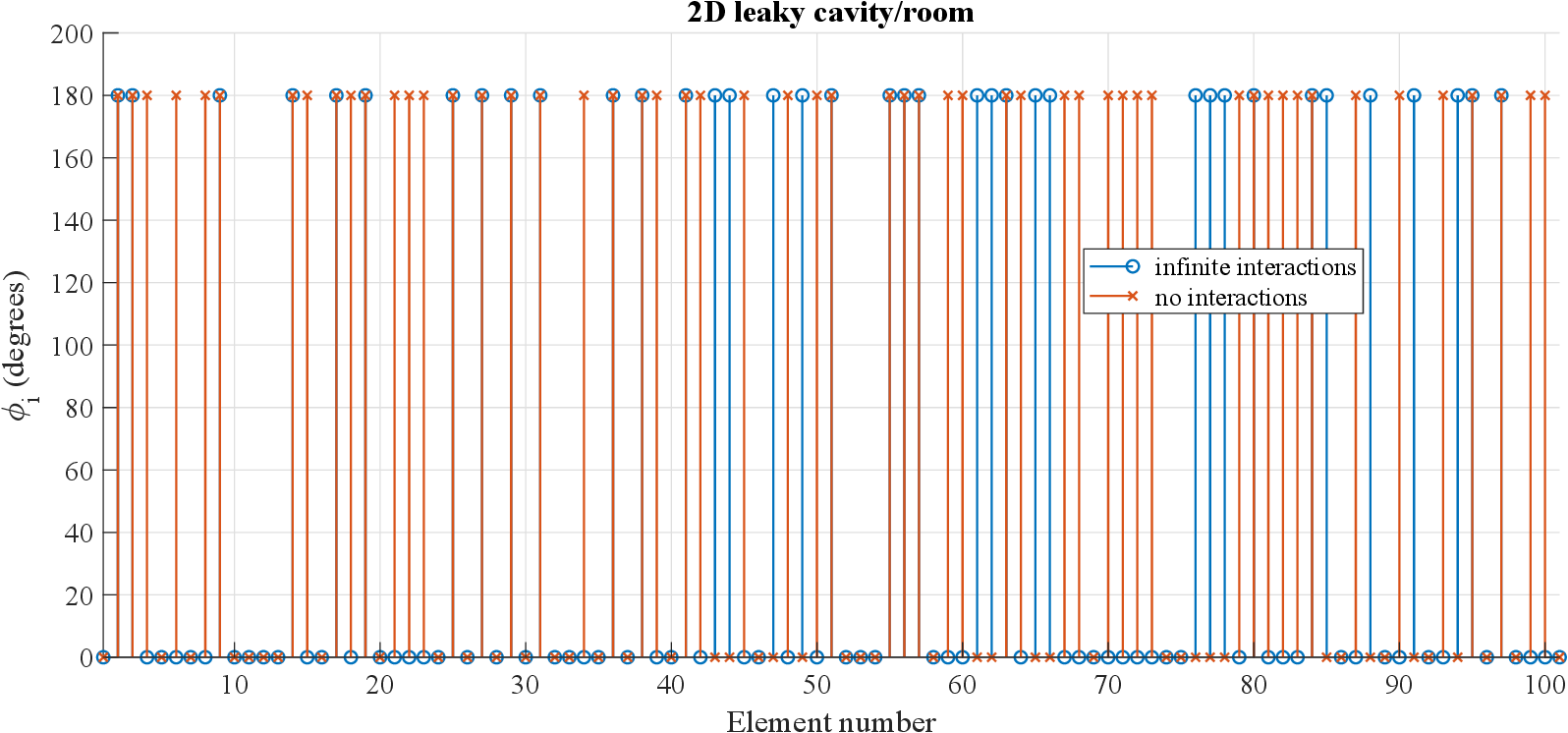} 
\caption{Phase set solutions, $\boldsymbol{\Phi}_{\infty}$ and $\boldsymbol{\Phi}_{0}$, for a RIS placed in a 2D leaky cavity/room.
\vspace{6mm}} \label{fig:angles-2}
\end{figure}

\begin{figure}[!ht]
\centering 
\includegraphics[width=1\textwidth]{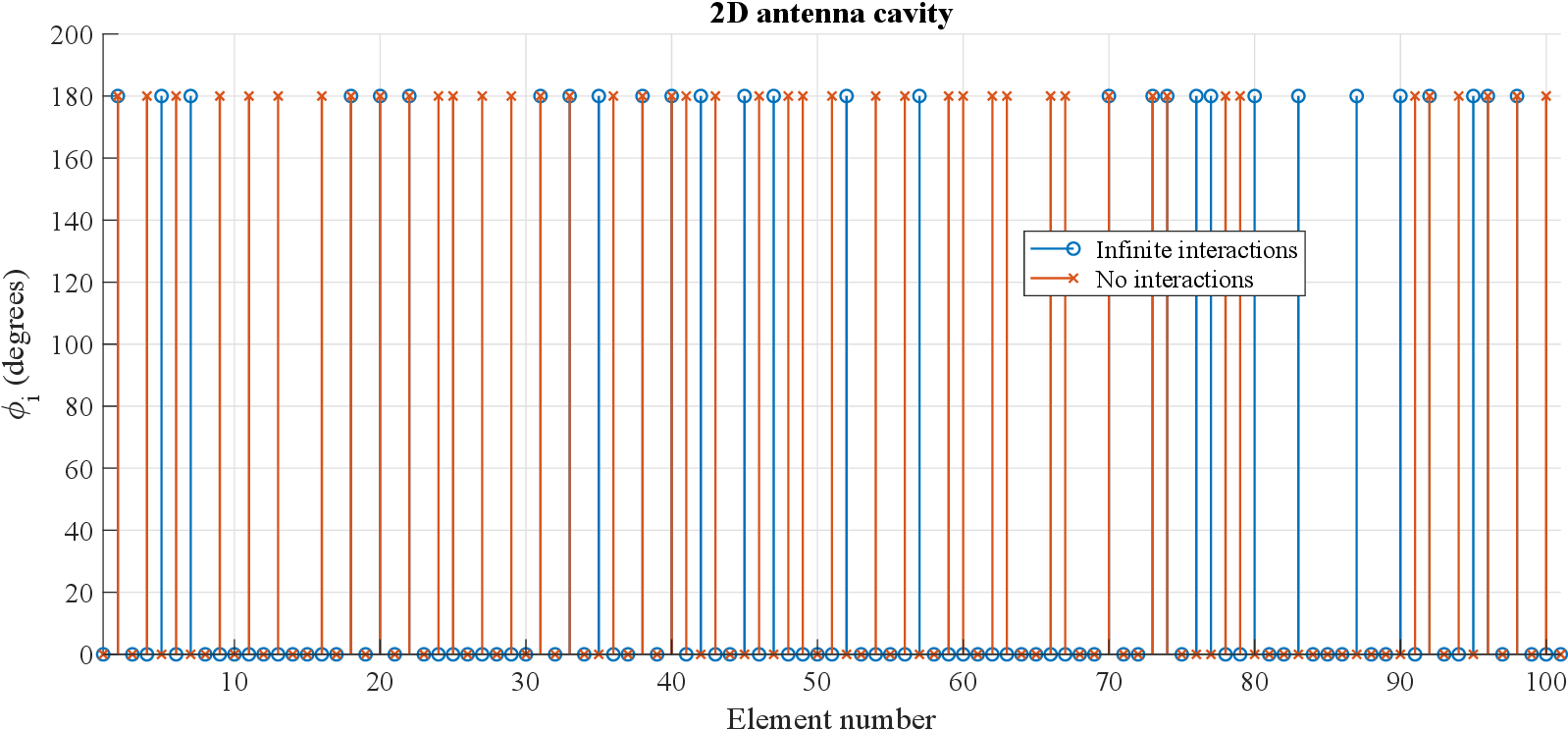}
\caption{Phase set solutions, $\boldsymbol{\Phi}_{\infty}$ and $\boldsymbol{\Phi}_{0}$, for a RIS placed in a 2D antenna cavity.
\vspace{1mm}} \label{fig:angles-3}
\end{figure}

In the \textbf{Section II} of the \textit{Main article}, the proposed topology optimization method is theoretically formulated. The application of this method in binary reconfigurable metasurfaces placed in a complex environment and with a goal of focusing at one point in space, will produce a set or a vector of phases with values $0$ or $\pi$. This phase set solution, if applied as a configuration to the tunable metasurface of the respective problem,  results to a focusing point at the expected position. Thus, the resulting phase values of each metasurface element for the 2D free space problem of \textbf{Section III.A} are displayed in Fig.\ref{fig:angles-1}, for taking into account infinite interactions between the elements, as well as for omitting them. We can observe that, due to the included binarization schemes discussed in \textbf{Section II.B}~\cite{christiansen2021inverse-s}, the phase output consists of only $0$ or $\pi$ values, according to the capabilities of the SMM unit cells~\cite{kaina2014hybridized-s}. The solutions for the 2D leaky cavity environment of \textbf{Section III.B} are depicted in Fig.\ref{fig:angles-2}, while the solutions for the 2D cavity antenna of \textbf{Section III.C} are shown in Fig.\ref{fig:angles-3}.

\newpage
\section{Application of the method of images for the 2D cavity antenna problem}
In this section, we semi-analytically calculate the values of $[H]$, $\textbf{H}_{F}$, $\textbf{H}_{S}$ and $H_{F\!S}$, as expressed in (7), for the cavity antenna problem depicted in Fig.5(a) of the \textit{Main article}, using the method of images.

First, let us consider a 2D unbounded, homogenous space, described by the cylindrical coordinate system $(\rho,\theta)$, with the vector to an observation point, $\boldsymbol{\rho} = \rho\boldsymbol{\hat{\rho}} =  \rho \left({\rm cos}\theta \mathbf{\hat{x}} +  {\rm sin}\theta \mathbf{\hat{y}}\right)$ and a unitary, and out-of-plane point current source placed at  $\boldsymbol{\rho}'$. The solution of this problem  is the two-dimensional Green’s function and is described by the following Helmholtz equation:
\begin{equation}\label{s-2d-green-function-free-1}
\left(\nabla^2 + k^2 \right)H(\boldsymbol{\rho},\boldsymbol{\rho}')\hat{\textbf{z}} = -\delta(\boldsymbol{\rho} - \boldsymbol{\rho}')\hat{\textbf{z}},
\end{equation}
\noindent where $k=\frac{\omega}{c}$ is the freespace wavenumber. Analytically, it writes:
\begin{equation}\label{s-2d-green-function-free-2}
H(\boldsymbol{\rho},\boldsymbol{\rho}') = G_{\rm 2d} (\|\boldsymbol{\rho}-\boldsymbol{\rho}'\|) =  -\frac{\mathrm{i}}{4}H^{(2)}_0(kR),
\end{equation}
where $H^{(2)}_0(.)$ denotes the zeroth-order Hankel function of the second kind and $R=\|\boldsymbol{\rho} - \boldsymbol{\rho}'\|$.
\begin{figure}[t]
\centering 
\includegraphics[width=1\textwidth]{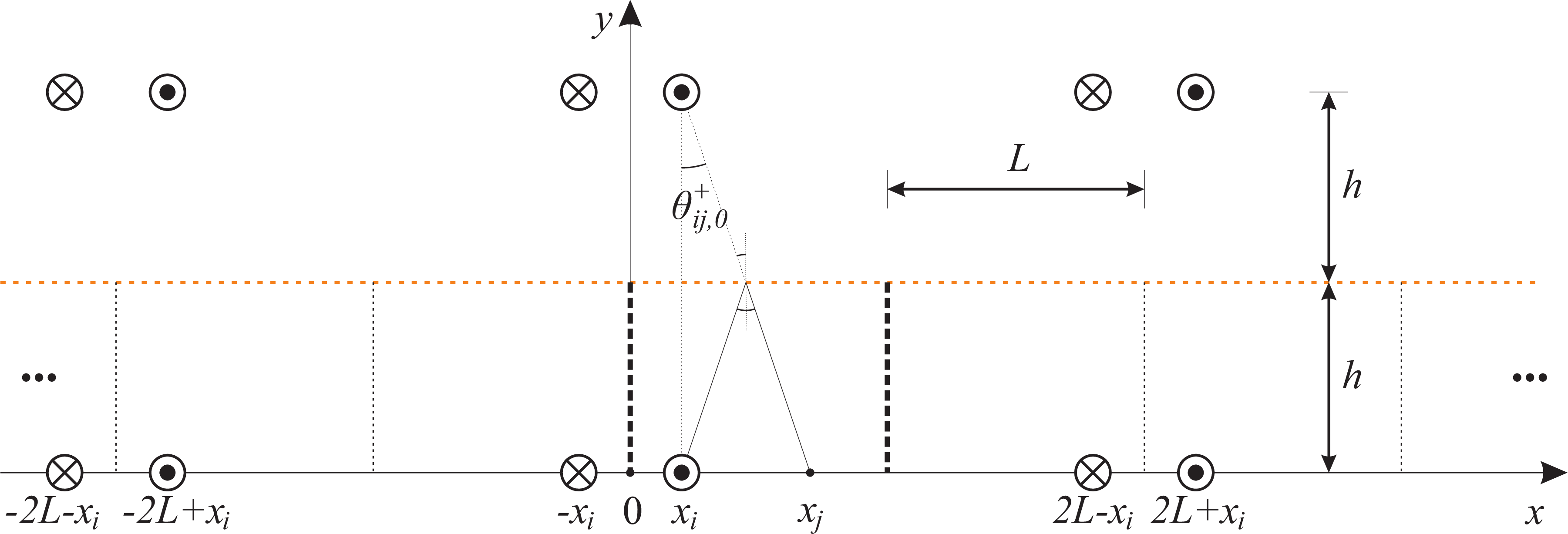} 
\caption{Application of the method of images for a single equivalent point source approximating an element of a reconfigurable metasurface inside the 2D cavity antenna of Fig.5(a) of the \textit{Main document} for the calculation of the interaction between elements. The bold dash black lines represent the removed PEC walls, while the dash orange line the removed partially reflective surface.
\vspace{0mm}} \label{fig:hij}
\end{figure}

Let us now focus on the problem of the leaky cavity antenna depicted in Fig.5(a) of \textbf{Section III.C}. First, it is assumed that the reflection and transmission coefficients at the partially reflective surface, or  $r$ and $t$, respectively, can be either retrieved by simulations or calculated, as a function of the angle of incidence.  It should be noted that the metasurface elements are placed  exactly at the bottom of the cavity and therefore there is no reflection on this wall.

We then begin from the calculation of the $H_{ij}$ values via the method of images~\cite{jackson1999classical-s,volakis2012integral-s}. 
In Fig.\ref{fig:hij}, the application of the method of images is illustrated. Initially, it is assumed that a point source is placed at $x=x_i$ and the goal is to calculate the interaction coefficient between the unit cell $i$ and the unit cell $j$ at $x=x_j$. Then, the PEC walls are removed and they are replaced by an infinite series of images of the source at $x=x_i$, with the remark that a PEC wall reverses the direction of the current of the subsequent image~\cite{volakis2012integral-s}. It is evident that the interaction coefficient $H_{ij}$ consists of two parts, the direct interaction along the $x$-axis and the interaction from the partially reflective surface above at $y=h$. Let us begin from the direct interaction part. The distance between the position of the images and $x=x_j$ is $d_{ij,m}^{\,\pm} = |x_j - (\pm x_i + 2mL)|$, where $m \in \mathbb{Z}$ denotes the number of images taken into account in the calculation, i.e. $2m+1$ images. A relatively large $m$ value will facilitate accurate results, without increasing the computational time of the overall algorithm; in this work, we use the value $m=30$, namely $61$ images. Therefore, the direct interaction between elements $i$ and $j$ is calculated as a summation of the 2D Green function values along the $x$-axis or,
\begin{equation}\label{s-Hij-d}
H^{\,\rm d}_{ij} = \sum_{-m}^m \Big\{G_{\rm 2d}(d^{\,+}_{ij,m}) - G_{\rm 2d}(d^{\,-}_{ij,m}) \Big\},\quad i \neq j    
\end{equation}\vspace{0mm}
For the calculation of the part related with the partially reflected surface above, we need to remove the surface and replace it again with point source images at $y=2h$, but this time their interaction is reduced by the reflection coefficient $r$. The reflection angle of a wave from a point source image impinging on the $j$ element is calculated as $\theta_{ij,m}^{\,\pm} = {\rm tan}^{-1}(d_{ij,m}^{\,\pm}/2h)$, while the distance of the images at $y=2h$ and $j$ is $q^{\,\pm}_{ij,m} = \sqrt{4h^2 + (d_{ij,m}^{\,\pm})^2}$, as deduced from Fig.\ref{fig:hij}. Thus, the reflection part of $H_{ij}$ is calculated as,
\begin{equation}\label{s-Hij-r}
H^{\,\rm r}_{ij} = \sum_{-m}^m \Big\{r(\theta^{\,+}_{ij,m})\,G_{\rm 2d}(q^{\,+}_{ij,m}) 
- r(\theta^{\,+}_{ij,m})\,G_{\rm 2d}(q^{\,-}_{ij,m}) \Big\}, \quad i \neq j.   
\end{equation}
Eventually, the full interaction between the elements $i$ and $j$ is derived from \eqref{s-Hij-d} and \eqref{s-Hij-r} as,
\begin{equation}\label{s-Hij}   
H_{ij} = H^{\,\rm d}_{ij} + H^{\,\rm r}_{ij}.
\end{equation}
The elements of the main diagonal of $[H]$ represent the self-interaction of each element $i$ originating from multiple reflections from the PEC walls and the partially reflecting surface and they are derived in a similar fashion as the $H_{ij}$ with an angle of incidence always zero, or $\theta_{ij,m}^{\,\pm} = 0$. Hence,
\begin{equation}\label{s-Hii} 
H_{ii} = r(0)\,G_{\rm 2d}(2h)  +
\sum_{-m}^m \Big\{G_{\rm 2d}(d^{\,+}_{ii,m}) - G_{\rm 2d}(d^{\,-}_{ii,m}) \Big\}.  
\end{equation}
\begin{figure}[t]
\centering 
\includegraphics[width=1\textwidth]{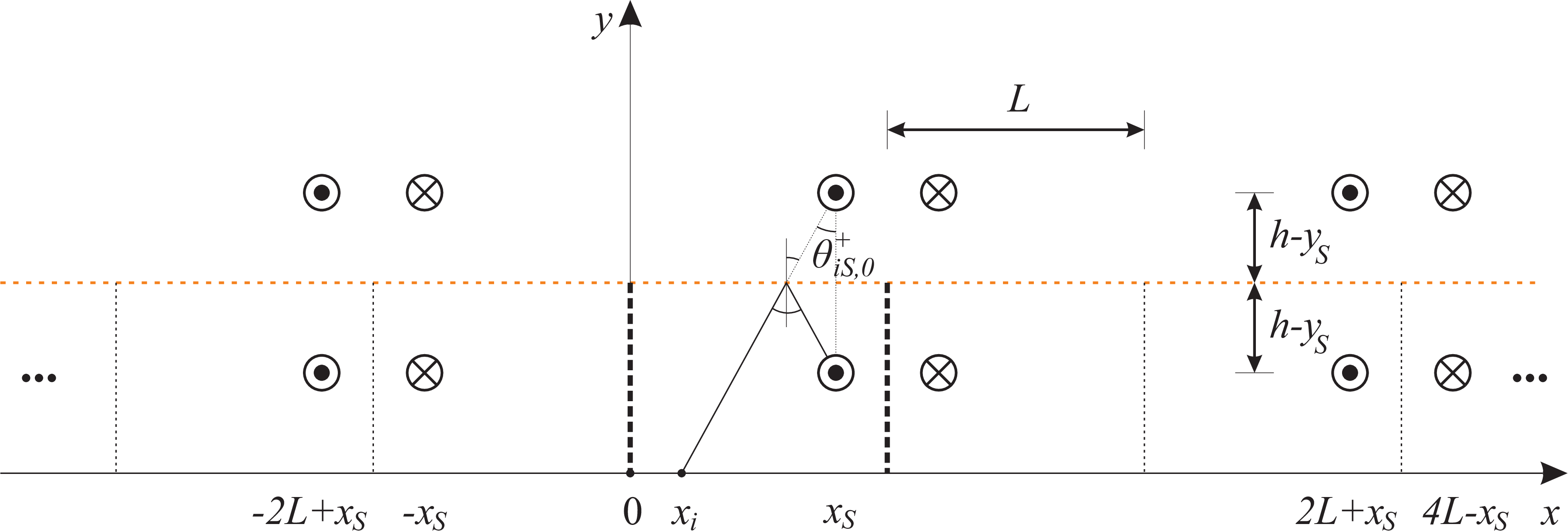} 
\caption{Application of the method of images for a point source feed inside the 2D cavity antenna of Fig.5(a) of the \textit{Main document} for the calculation of the interaction between the source and a metasurface element. 
\vspace{0mm}} \label{fig:his}
\end{figure}
The interaction between the main point source at $S(x_{S},y_{S})$, represented by the vector elements $H_{Si}$, is similarly calculated with method of images, as depicted in Fig.\ref{fig:his}. The direct interaction, here, is the one from the images below the partially reflective surface, or at $y<h$. The distance of these source images and the unit-cell $i$ is $q^{\,\pm}_{1,i,m} = \sqrt{y_{S}^2 + (d_{Si,m}^{\,\pm})^2}$ with $d_{Si,m}^{\,\pm} = |2mL \pm x_{S} - x_i|$. Thus, the direct interaction coefficient is derived as,
\begin{equation}\label{s-His-d}
H^{\,\rm d}_{Si} = \sum_{-m}^m \Big\{G_{\rm 2d}(q^{\,+}_{1,i,m}) - G_{\rm 2d}(q^{\,-}_{1,i,m}) \Big\}.  
\end{equation}
Similarly to the previous case of $H^{\rm r}_{ij}$ in \eqref{s-Hij-r}, originating from reflections from the partially reflective surface, the distance between the unit-cell $i$ and the images of the source at $y = 2h - y_{S}$  is $q^{\,\pm}_{2,i,m} = \sqrt{(2h^2-y_{S})^2 + (d_{Si,m}^{\,\pm})^2}$, while the angle of incidence on the partially reflective surface for a wave from each source image to the unit-cell $i$ is $\theta_{Si,m}^{\,\pm} = {\rm tan}^{-1}(d_{Si,m}^{\,\pm}/(2h-y_{S}))$. Thus, the second part of $H_{Si}$ is calculated as,
\begin{equation}\label{s-His-r}
H^{\,\rm r}_{Si} = \sum_{-m}^m \Big\{r(\theta^{\,+}_{Si,m})\,G_{\rm 2d}(q^{\,+}_{2,i,m}) 
- r(\theta^{\,+}_{Si,m})\,G_{\rm 2d}(q^{\,-}_{2,i,m}) \Big\},
\end{equation}
and the total $H_{Si}$ from~\eqref{s-His-d} and~\eqref{s-His-r} is
\begin{equation}\label{s-His}
H_{Si} = H^{\,\rm d}_{Si} + H^{\,\rm r}_{Si}.
\end{equation}

\begin{figure}[t]
\centering 
\includegraphics[width=1\textwidth]{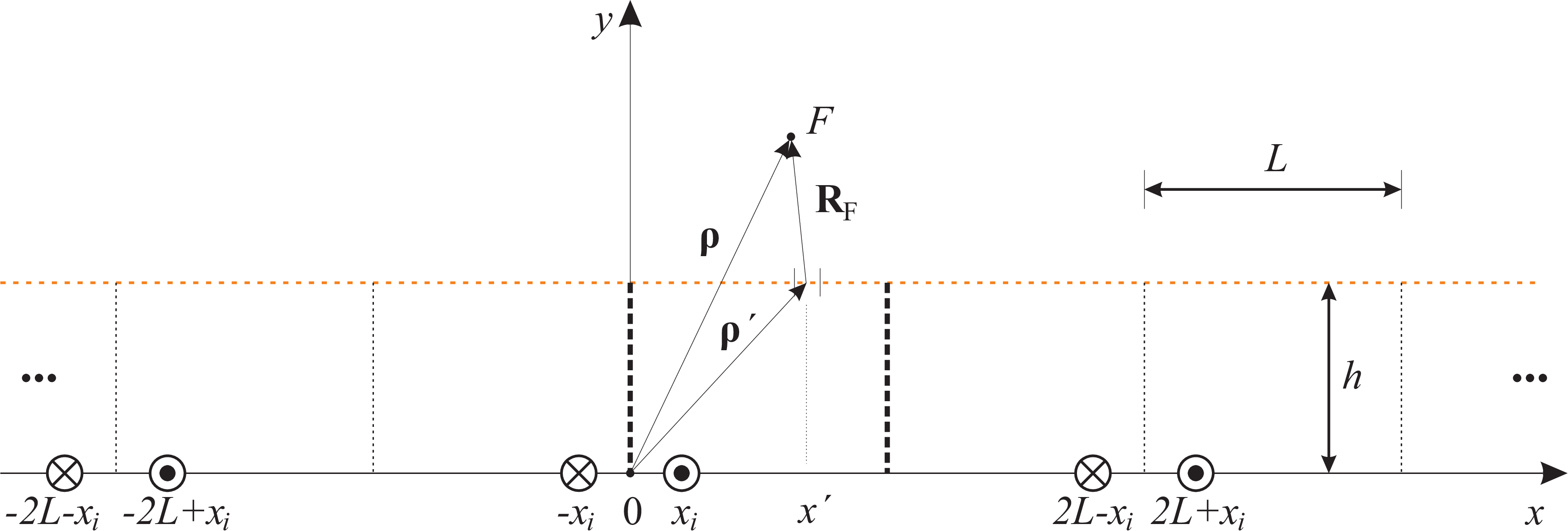} 
\caption{Application of the method of images for a single equivalent point source approximating an element of a reconfigurable metasurface inside the 2D cavity antenna of Fig.5(a) of the \textit{Main document} for the calculation of the fields at the point $F$ via the Kirchhoff integral.
\vspace{0mm}} \label{fig:Eix}
\end{figure}
The calculation of the remaining values of $H_{Fi}$ and $H_{F\!S}$ is less straightforward than the ones above and requires the accurate calculation of the fields outside the cavity. This is possible by utilizing the \textit{Kirchhoff integral}~\cite{jackson1999classical-s}, which states that if inside a closed volume $V$ with a closed surface $S'$ there are sources, then, the E-field at a point $\boldsymbol{\rho}$ outside can be calculated as,
\begin{equation}\label{s-kirchhoff-int-1}
E(\boldsymbol{\rho}) = \oiint_{S'} E(\boldsymbol{\rho}')\, \bigg(\mathbf{\hat{n}'}\cdot\nabla'G_{\rm 2d}(\boldsymbol{\rho},\boldsymbol{\rho}')\bigg) 
-\, G_{\rm 2d}(\boldsymbol{\rho},\boldsymbol{\rho}')\,\bigg(\mathbf{\hat{n}'}\cdot\nabla'E(\boldsymbol{\rho}') \bigg)\, dS',    
\end{equation}
where $\mathbf{\hat{n}'}$ is the unit vector normal to $S'$ and directed \textit{into} $V$. The~\eqref{s-kirchhoff-int-1} basically considers all infinitesimally small sections of the surface of the closed cavity as emitting point sources according to the Huygens principle~\cite{jackson1999classical-s}. If we apply~\eqref{s-kirchhoff-int-1} in the 2D cavity antenna problem, as depicted on Fig.5(a) of \textbf{Section III.C} the \textit{Main article}, no electric field can be found at the outside PEC wall surfaces and below the metasurface elements. Therefore, only the scattered field from the partially reflective surface contributes to the total field outside the cavity and~\eqref{s-kirchhoff-int-1} is simplified to:
\begin{equation}\label{s-kirchhoff-int-2}
E(\boldsymbol{\rho}_{F}) = \int_0^L E(\boldsymbol{\rho}')\, \bigg(\mathbf{\hat{y}}\cdot\nabla'G_{\rm 2d}\big(\|\boldsymbol{\rho}_{F}-\boldsymbol{\rho}'\|\big)\bigg) 
+\, G_{\rm 2d}\big(\|\boldsymbol{\rho}_{F}-\boldsymbol{\rho}'\|\big)\,\bigg(\mathbf{\hat{y}}\cdot\nabla'E(\boldsymbol{\rho}')\bigg)\, dx',    
\end{equation}
where the vector $\boldsymbol{\rho}'$ represents the position of each infinitely small piece of the integral, while $\boldsymbol{\rho}_{ F}$ represents the position of the point where the field is to be calculated, herein, the focusing point. Hence, $\|\boldsymbol{\rho}_{F}-\boldsymbol{\rho}'\| = R_{F} = \sqrt{(x_{F} - x')^2 + (y_{F} - h)^2}$. The parts of the integral involving Green functions and their derivatives can be directly substituted via \eqref{s-2d-green-function-free-2} or calculated as, 
\begin{equation}\label{s-kirchhoff-int-left}
\mathbf{\hat{y}}\cdot\nabla'G_{\rm 2d}(R_{F}) 
=\frac{|y_{F} - h|}{\sqrt{(x_{\rm F} - x')^2 - (y_{F} - h)^2}}\, \Bigg[\frac{\mathrm{i}k}{4}H^{(2)}_1(kR_{F})\Bigg].    
\end{equation}

Then, the problem of calculating $H_{Fi}$ and $H_{FS}$ essentially becomes a problem of calculating the E-fields and their derivatives on each point of the upper part of partially reflected surface, after placing a current point source at the metasurface elements positions or at $S$ and using the method of images, and, subsequently, a problem of calculating via \eqref{s-kirchhoff-int-2} the E-field at $F$. The E-fields on the upper part of partially reflected surface could be approximately calculated analytically, through the Fresnel-Kirchhoff diffraction formula \cite{jackson1999classical-s,hecht2012optics-s}, because in the antenna cavity problem of \textbf{Section III.C} we are using a 2D aperture array. Nevertheless, in this work, the necessary E-fields for \eqref{s-kirchhoff-int-2} are derived by calculating the field on the lower part using image theory and Green functions and, then, by multiplying it with the extracted transmission coefficient of the partially reflective surface \cite{comsol-s}. It should be noted that the transmission coefficient is extracted here on ports at a large distance and, thus, all evanescent modes are omitted. Therefore, this is a far field approximation and it is expected to produce inaccurate results close to the partially reflective surface under study. Nevertheless, this procedure is valid for antenna applications, like the presented cavity antenna in \textbf{Section III.C}. 
\begin{figure}[t]
\centering 
\includegraphics[width=1\textwidth]{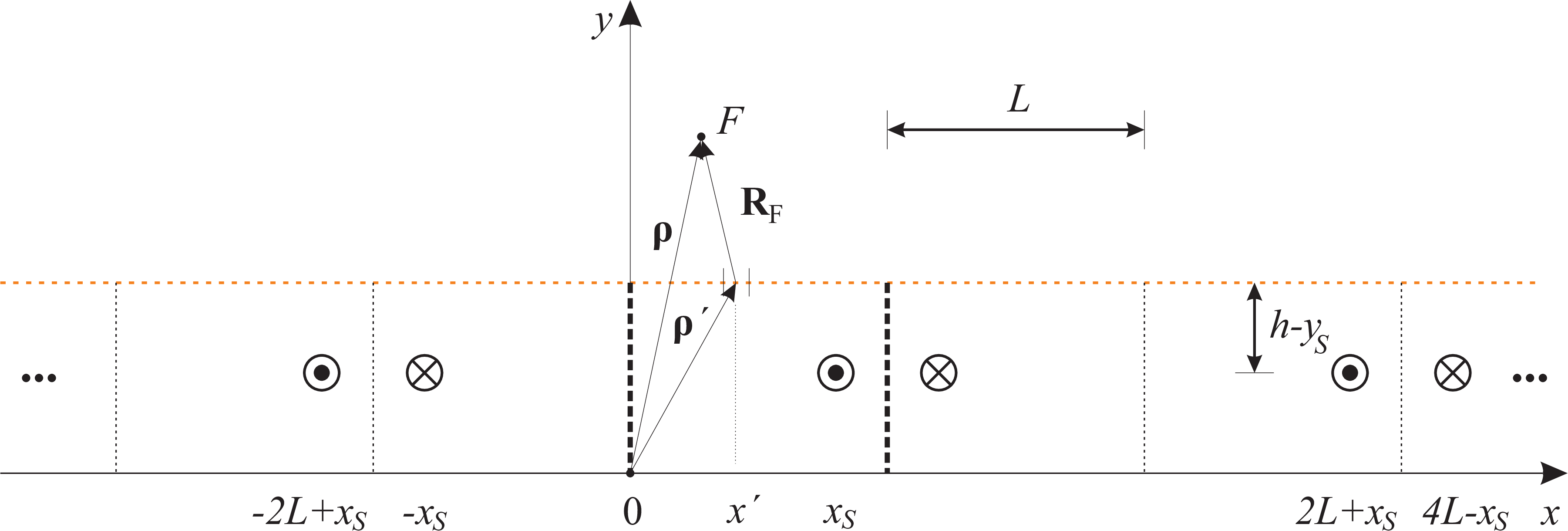} 
\caption{Application of the method of images for or a point source feed inside the 2D cavity antenna of Fig.5(a) of the \textit{Main document} for the calculation of the fields at the point $F$ via the Kirchhoff integral.
\vspace{0mm}} \label{fig:Eis}
\end{figure}
We start from the calculation of E-fields on the surface points at $y=h$ for a point source $I_i$ placed at each unit cell $i$. In Fig.\ref{fig:Eix}, the application of the method  of images is illustrated. The distance between each image of the unit cell $i$ and the point $x'$ is calculated as $u_{i,m}^{\,\pm} = \sqrt{h^2 + (x' + 2mL \pm x_{i})^2}$, while the angle of incidence of the wave from each image to $x'$ as $\theta_{{\rm S}x',m}^{\,\pm} = {\rm tan}^{-1}(u_{{\rm S},m}^{\,\pm}/(y_{\rm S} - h))$.
Therefore, for each metasurface element $i$ it is derived that,
\begin{subequations}\label{s-Ei}  
\begin{equation}
E_{\,ix'} = \sum_{-m}^m \Big\{t(\theta^{\,+}_{ix',m})\,G_{\rm 2d}(u^{\,+}_{i,m}) 
-t(\theta^{\,-}_{ix',m})\,G_{\rm 2d}(u^{\,-}_{i,m}) \Big\}\,(-\mathrm{i}\omega\mu I_i),  
\end{equation}
\begin{equation}
\mathbf{\hat{y}}\cdot\nabla'E_{\,ix'} = \frac{-\mathrm{i}k}{4}\sum_{-m}^m\Big\{\frac{h}{u^{\,+}_{i,m}}\,
t(\theta^{\,+}_{ix',m})\,H^{(2)}_1(k\,u^{\,+}_{i,m}) - \frac{h}{u^{\,-}_{i,m}}\,
t(\theta^{\,-}_{ix',m})\,H^{(2)}_1(k\,u^{\,-}_{i,m})
\Big\}\,(-\mathrm{i}\omega\mu I_i),
\end{equation}
\end{subequations}

We will calculate in a similar fashion the E-field on the partially reflective surface from the antenna cavity main source $S$, as illustrated in Fig.\ref{fig:Eis}. The distance between each image of the source and the point $x'$ on the surface is calculated as $u_{{S},m}^{\,\pm} = \sqrt{(y_{S} - h)^2 + (x' + 2mL \pm x_{S})^2}$, while the angle of incidence of the wave from each image to $x'$ as $\theta_{{S}x',m}^{\,\pm} = {\rm tan}^{-1}(u_{{S},m}^{\,\pm}/(y_{S} - h))$. Thus, for the source $S$ it is calculated that,
\begin{subequations}\label{s-Es}  
\begin{equation}
E_{\,{S}x'} = \sum_{-m}^m \Big\{t(\theta^{\,+}_{{S}x',m})\,G_{\rm 2d}(u^{\,+}_{{S},m}) \,-
t(\theta^{\,-}_{{S}x',m})\,G_{\rm 2d}(u^{\,-}_{{S},m}) \Big\}\,(-\mathrm{i}\omega\mu I_i),  
\end{equation}
\begin{equation}
\mathbf{\hat{y}}\cdot\nabla'E_{\,{S}x'} = \frac{-\mathrm{i}k}{4}\sum_{-m}^m\Big\{\frac{h}{u^{\,+}_{{S},m}}
\,t(\theta^{\,+}_{{S}x',m})\,H^{(2)}_1(ku^{\,+}_{{S},m}) 
 -\frac{h}{u^{\,-}_{{\rm S},m}}\,
t(\theta^{\,-}_{{\rm S}x',m})\,H^{(2)}_1(ku^{\,-}_{{S},m})
\Big\}\,(-\mathrm{i}\omega\mu I_i),
\end{equation}
\end{subequations}

Finally, inserting \eqref{s-2d-green-function-free-2}, \eqref{s-kirchhoff-int-left} and \eqref{s-Ei} into \eqref{s-kirchhoff-int-2} produces $E_{Fi}$, which in turn leads to the calculation of the required vector element via $H_{Fi} = E_{Fi}/(-\mathrm{i}\omega\mu I_i)$. Similarly, inserting \eqref{s-2d-green-function-free-2}, \eqref{s-kirchhoff-int-left} and \eqref{s-Es} into \eqref{s-kirchhoff-int-2} will give $E_{{FS}}$ and, eventually, $H_{FS} = E_{{FS}}/(-\mathrm{i}\omega\mu I_i)$, thus, providing the last value required for (8) in the \textit{Main article}.

\begin{figure}[ht]
\centering 
\includegraphics[width=0.475\textwidth]{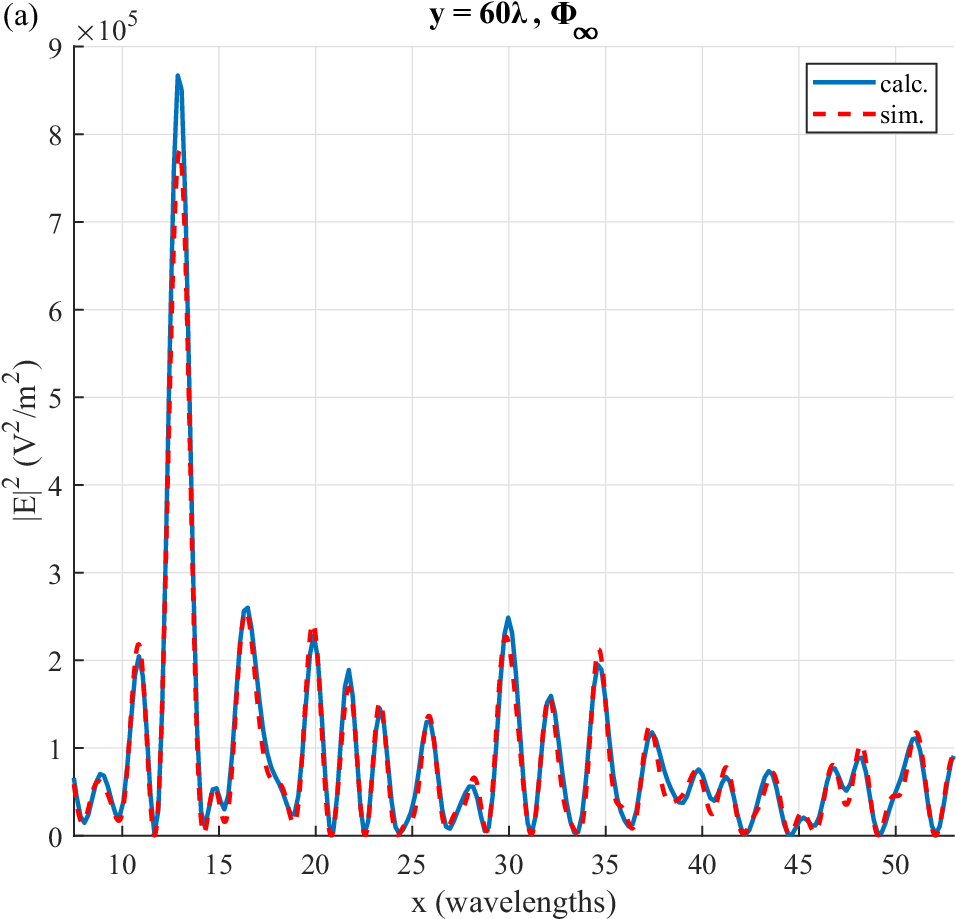} 
\includegraphics[width=0.49\textwidth]{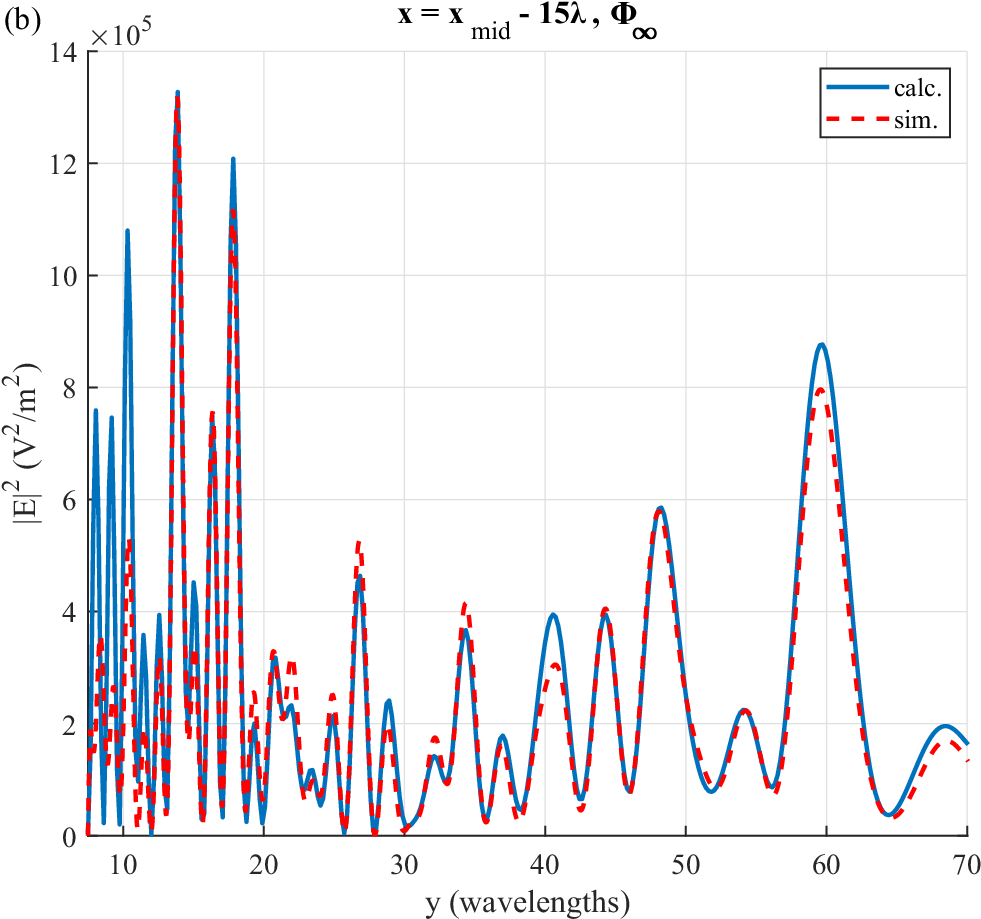} 
\caption{Comparison of the analytical and simulated intensities around the focusing point $F$ for (a) $y=y_{F}$ and (b) $x=x_{F}$.
\vspace{0mm}} \label{fig:validation}
\end{figure}

Let us now validate model presented in this section via numerical simulations. In \textbf{Section III.C} of the \textit{Main article}, we demonstrated the resulting focus at $F(x_{\rm F},y_{\rm F})$ for infinite interactions in Figs.5(b). We will now calculate the fields on $y = y_{F}$ and $x = x_{F}$ using the formulas developed in this Section, the phase solutions of the topology optimization for this case and (8), and we will compare the resulting intensities with the ones after a numerical simulation \cite{comsol-s} of the 2D cavity antenna setup using the same phase solutions. The results are comparatively demonstrated in \ref{fig:validation}, and they show an very good agreement in the far field, or sufficiently away from the partially reflective surface at $y=7.5\lambda$, thus, certifying the validity of the proposed semi-analytical model for a 2D antenna cavity. 

\newpage

\bibliographystyle{ieeetr}

\end{document}